\documentclass[8.5pt,twoside,twocolumn]{article}
\usepackage[super,sort&compress,comma]{natbib} 
 \usepackage[version=3]{mhchem}
\usepackage{times}
\usepackage{sectsty}
\usepackage{balance} 

\usepackage{graphicx} 
\usepackage{lastpage}
\usepackage[format=plain,justification=raggedright,singlelinecheck=false,font=small,labelfont=bf,labelsep=space]{caption} 
\usepackage{fancyhdr}                      
\usepackage[usenames,dvipsnames]{xcolor}

\begin{document}

\thispagestyle{plain}
\fancypagestyle{plain}{
\renewcommand{\headrulewidth}{1pt}}
\renewcommand{\thefootnote}{\fnsymbol{footnote}}
\renewcommand\footnoterule{\vspace*{1pt}%
\hrule width 3.4in height 0.4pt \vspace*{5pt}} 
\setcounter{secnumdepth}{5}

\makeatletter 
\def\subsubsection{\@startsection{subsubsection}{3}{10pt}{-1.25ex plus -1ex minus -.1ex}{0ex plus 0ex}{\normalsize\bf}} 
\def\paragraph{\@startsection{paragraph}{4}{10pt}{-1.25ex plus -1ex minus -.1ex}{0ex plus 0ex}{\normalsize\textit}} 
\renewcommand\@biblabel[1]{#1}            
\renewcommand\@makefntext[1]%
{\noindent\makebox[0pt][r]{\@thefnmark\,}#1}
\makeatother 
\renewcommand{\figurename}{\small{Fig.}~}
\sectionfont{\large}
\subsectionfont{\normalsize} 

\fancyfoot{}
\fancyfoot[RO]{\footnotesize{\sffamily{1--\pageref{LastPage} ~\textbar  \hspace{2pt}\thepage}}}
\fancyfoot[LE]{\footnotesize{\sffamily{\thepage~\textbar\hspace{3.45cm} 1--\pageref{LastPage}}}}
\fancyhead{}
\renewcommand{\headrulewidth}{1pt} 
\renewcommand{\footrulewidth}{1pt}
\setlength{\arrayrulewidth}{1pt}
\setlength{\columnsep}{6.5mm}
\setlength\bibsep{1pt}

\twocolumn[
  \begin{@twocolumnfalse}
\noindent\LARGE{\textbf{Tuning the Ferroelectric Polarization in $AA^{\prime}$MnWO$_6$ Double Perovskites through $A$ Cation Substitution}}
\vspace{0.6cm}

\noindent\large{\textbf{Joshua Young,$^\textit{$^{a}$}$ Alessandro Stroppa,\textit{$^{b}$} Silvia Picozzi\textit{$^{b}$}, and James M.\ Rondinelli,\textit{$^{c,\ast}$}}}\vspace{0.5cm}

\noindent\textit{\small{\textbf{Received Xth XXXXXXXXXX 20XX, Accepted Xth XXXXXXXXX 20XX\newline
First published on the web Xth XXXXXXXXXX 200X}}}

\noindent \textbf{\small{DOI: 10.1039/b000000x}}
\vspace{0.6cm}

\noindent \normalsize{Recent experimental and theoretical work has shown that the double perovskite NaLaMnWO$_6$ exhibits
antiferromagnetic ordering owing to the Mn $d$ states, and computational studies further predict it to exhibit a spontaneous electric polarization due to an improper mechanism for ferroelectricity [King \textit{et al., Phys.\ Rev.\ B}, 2009, \textbf{79}, 224428; Fukushima \textit{et al., Phys.\ Chem.\ Chem.\ Phys.}, 2011, \textbf{13}, 12186], which make it a candidate multiferroic material.  
Using first-principles density functional calculations, we investigate nine isostructural and isovalent  $AA^{\prime}$MnWO$_6$ double perovskites ($A$=Na, K, and Rb; $A^{\prime}$=La, Nd, and Y) with the aim of articulating crystal-chemistry guidelines describing how to enhance the magnitude of the electric polarization through chemical substitution of the $A$-site while retaining long-range magnetic order.
We find that the electric polarization can be enhanced by up to 150\% in compounds which maximize the difference in the ionic size of the $A$ and $A^{\prime}$ cations.
By examining the tolerance factors, bond valences, and structural distortions (described by symmetry-adapted modes) of the nine compounds, we identify the atomic scale features that are strongly correlated with the ionic and electronic contributions to the electric polarization.
We also find that each compound exhibits a purely electronic remnant polarization, even in the absence of a displacive polar mode.
The analysis and design strategies presented here can be further extended to additional members of this family ($B$=Fe, Co, etc.), and the improper ferroelectric nature of the mechanism allows for the decoupling of magnetic and ferroelectric properties and the targeted design of novel multiferroics.}
\vspace{0.5cm}
 \end{@twocolumnfalse}
  ]

\section{Introduction}
\footnotetext{\dag~Electronic Supplementary Information (ESI) available: Full crystallographic structure for the $AA^{\prime}$MnWO$_6$ compounds explored. See DOI: 10.1039/b000000x/}


\footnotetext{\textit{$^{a}$Department of Materials Science and Engineering, Drexel University, Philadelphia PA, 19104, USA. Email: jy346@drexel.edu}}
\footnotetext{\textit{$^{b}$Consiglio Nazionale delle Ricerche - CNR-SPIN, L'Aquila, Italy}}
\footnotetext{\textit{$^{c}$Department of Materials Science and Engineering, Northwestern University, Evanston, IL, 60208, USA}}



Magnetic ferroelectrics have received renewed interest over the past decade for their considerable scientific and technological promise.\cite{Eerentstein/Mathur/Scott:2006,Gajek_etal:2007,Nan_etal:2008}
Unfortunately, the mechanisms generating electric and magnetic polarizations are often mutually exclusive,\cite{Hill:2000}
making these types of materials challenging to find.
One key to accelerating their discovery lies in finding a general prescription for inducing both collective ferroic orders. 
One of the most successful approaches involves selecting compounds which display magnetic ordering, and then generating a spontaneous polarization through cooperative polar distortions of non-magnetic cations.\cite{Varga_Fennie_etal:2009,Ghosh_etal:2014,Li_Stephens_etal:2014,Lu/Xiang:2014}
To this end, recent investigations have focused on transition metal $AB$O$_3$ perovskite oxides, whose wide array of possible chemistries and physical properties make them ideal candidates for multiferroic discovery.
The typical mechanism for the appearance of a spontaneous electric polarization in ferroelectric perovskites is due to hybridization between the oxygen $2p$-states and the $A$ or $B$ metal cations through the second-order Jahn-Teller (SOJT) effect, resulting in cooperative polar displacements.\cite{Kunz/Brown:1995,Halasyamani/Poeppelmeier:1998,Bersuker:1995}
However, the fact that the SOJT effect is most energetically favorable for cations with either a stereoactive lone pair or a $d^0$ electronic configuration\cite{Burdett:1981} leads to competitions between centric and noncentrosymmetric structures and magnetic moment formation.\cite{Hill:2000}
While recent work has shown that it is possible for other $d^n$ configurations to exhibit the SOJT effect,\cite{Barone/Picozzi_CaMnO3:2011} the stringent criteria ($n$ = 3-7, and only specific spin states) limits the possible chemistries of new ferroelectric materials.\cite{Bersuker:2012}

Recently, an alternative route for the design of multiferroics has emerged with the identification of the so-called `hybrid improper' ferroelectric mechanism.\cite{Bousquet_etal:2008}
Here, the spontaneous electric polarization is produced as a byproduct of two \textit{non-polar} lattice instabilities ($Q_1$ and $Q_2$) which couple to a polar mode ($P$) through an anharmonic interaction in the free energy expansion of the form
$\gamma Q_1Q_2P$,\cite{Benedek/Fennie:2011}
where $\gamma$ is the coupling constant.
In systems consisting of extended metal-oxygen networks, the non-polar modes are often rotations of polyhedral building blocks.\cite{Martin/Chu/Ramesh:2010}
%
This mechanism was first proposed in the Aurivillius compound SrBi$_2$Nb$_2$O$_9$,\cite{PerezMatoHIF:2004} but these types of distortions are also ubiquitous\cite{Glazer:1972,Woodward1:1997,Woodward2:1997} and able to generate ferroelectricity in a wide variety of other solid state oxides, such as Ruddlesden-Popper phases,\cite{Benedek/Fennie:2011,PhysRevLett.112.187602} Dion-Jacobson phases,\cite{Benedek:2014} and perovskites (the focus of this work).\cite{Bousquet/Ghosez_et_al:2008,PhysRevB.89.174101,IniguezGhosez_etal:2013}
However, these modes are not restricted to only rotations; they can manifest themselves as any number of atomic displacement patterns, such as pseudo-rotations or Jahn-Teller distortions.\cite{Stroppa_CuMOF:2011,StroppaMOF:2013}
Because this mechanism relies on a `geometric' effect (\textit{e.g.} displacements of ions due to bond-coordination preferences and packing), in contrast to a chemical bonding or `electronic' mechanisms (such as the SOJT effect),\cite{Picozzi/Stroppa_MF:2012} there are many more potential routes for purposefully inducing and enhancing ferroelectricity by judicious selection of cation sizes. 

Such avenues have been thoroughly explored in cation ordered $AB$O$_3$ perovskite oxides,  \cite{Bousquet/Ghosez_et_al:2008,PhysRevB.89.174101,IniguezGhosez_etal:2013,Rondinelli/Fennie:2012,Mulder/Rondinelli/Fennie:2013,Young/Rondinelli:2013,StroppaMOF:2013,Benedek/Mulder/Fennie:2012}
because the nominally non-polar lattice modes that describe the in-phase ($Q_1$) 
and out-of-phase ($Q_2$) octahedral rotations common in orthorhombic perovskites will couple and may induce a polar lattice mode through the anharmonic lattice interaction.\cite{Bellaiche/Iniguez:2013,Bellaiche/Iniguez_etal:2014}
If the $A$ cations are ordered to create [$AB$O$_3$]$_1$/[$A^{\prime}B$O$_3$]$_1$ superlattices along the [001] direction, the polar mode is largely characterized by anti-polar displacements of $A$ cations with unequal amplitude that occur in the  $A$O and $A^{\prime}$O (001) planes. %
The different $A$ and $A^\prime$ ionic sizes and bond valence preferences produce the net electric polarization. 
Because the polarization results from the $A$-site displacements, the $B$-sites can be occupied by magnetic cations without reducing the tendency to ferroelectricity.
Such a prescription thus allows for the engineering of both electric and magnetic polarizations (multiferroism) in the same material.

\begin{figure}[t]
\centering
\includegraphics[width=0.8\columnwidth,clip]{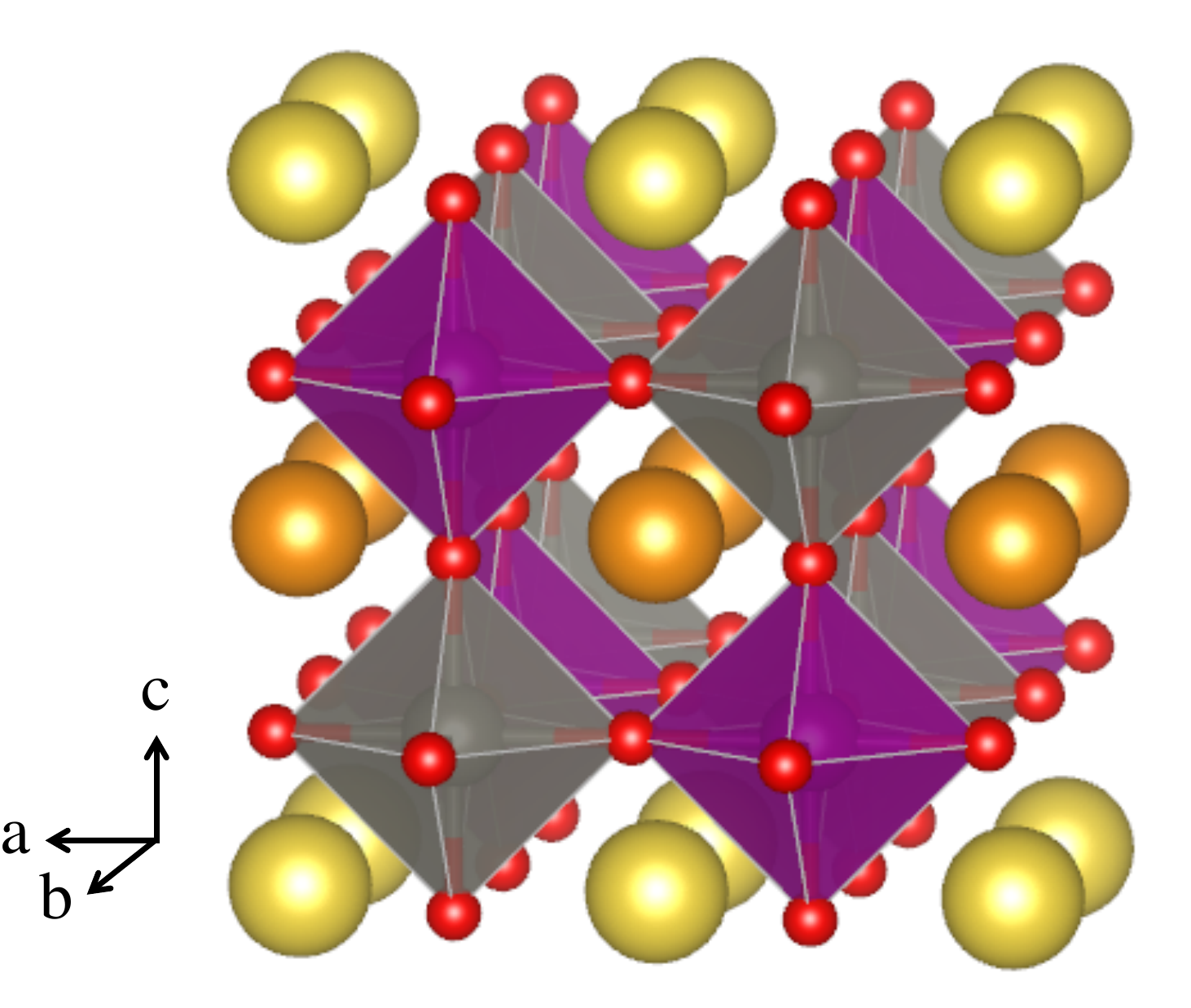}\vspace{-1\baselineskip}
\caption{The crystal structure of an $AA^{\prime}BB^{\prime}$O$_6$ double perovskite oxide with ordering of the $A$-site (orange and yellow) and $B$-site (purple and grey) cations along the [001] and [111] crystallographic axes, respectively. This high symmetry structure shown here exhibits space group $P4/nmm$ (\#129).
}
\label{fig:P4nmm}
\end{figure}

Here, we focus on the family of $AA^{\prime}BB^{\prime}$O$_6$ double perovskites, which 
thermodynamically favor simultaneous layering of the $A$-sites and rock salt patterning of the $B$-sites (Figure \ref{fig:P4nmm}).\cite{Graham/Woodward:2010}
This type of chemical ordering, in combination with the common $a^-a^-c^+$ octahedral rotation pattern, is also capable of lifting inversion symmetry through a hybrid improper mechanism.
Although many double perovskites have been experimentally made with a wide variety of different $B$ cations, including Mg, Mn, Co, Ni, Ti, Fe, W, Sc, Te, and Nb,\cite{Lopez/Veiga/Pico:1994,King/WaymanWoodward:2009,Knapp/Woodward:2006,King/Woodward_etal:2007,Arillo/Gomez_etal:1997,Arillo/Gomez_etal_2:1997} we choose the series of $AA^{\prime}$MnWO$_6$ compounds for which the Mn$^{2+}$ atoms are known to order magnetically, and the W$^{6+}$ cations provide the large valance state difference needed for rock salt ordering.\cite{Graham/Woodward:2010}

The NaLaMnWO$_6$ member of this family has been synthesized experimentally, and exhibits a polar $P2_1$ ground state at room temperature in addition to a G-type antiferromagnetic ordering of the Mn$^{2+}$ atoms below 10 K; the N\'{e}el temperature is low due to the fact that the $B$-site rock salt ordering separates the Mn atoms, and the superexchange interaction occurs via an Mn-O-W-O-Mn pathway with non-magnetic tungsten.\cite{King/Wills/Woodward:2009}
Additionally, a large electric polarization of 16 $\mu$C/cm$^2$ arising via a hybrid improper mechanism is predicted from density functional theory calculations.\cite{Fukushima_etal:2011}
A recent experimental study, however, found zero spontaneous polarization in samples of NaLaMnWO$_6$ and NaNdMnWO$_6$.\cite{De_Kim_etal:2014}
This discrepancy could be attributed to the polycrystalline samples and difficulties in poling 
sintered ceramic pellets.

In this work, we investigate a series of nine iso-structural compounds with a combination of alkali earth metals on the $A$-site ($A$=Na, K, Rb) and rare-earth cations on the $A^\prime$-site ($A^\prime$=La, Nd, Y) using first-principles density functional theory calculations.
We find all materials in this family display a spontaneous polarization, which is enhanced by up to 150\% in those chemistries which maximize the $A$-$A^{\prime}$ cation size differential.
In order to understand the origin and requirements for the presence of this net dipole, we examine a variety of structural descriptors, including tolerance factors, displacive mode amplitudes, bond valence sums, and layer decomposed dipoles.
We find that while the size of the rare earth cation is the largest factor controlling the magnitude of the polarization, consideration of the alkali metal is needed to fully distinguish the compounds with the largest polarization in a single lanthanide family. 
With this understanding, we build a predictive model based on crystal-chemistry factors to describe other double perovskites, locate new compositions, as well as determine how to optimize the electric polarization and intrinsic switching barrier.
Finally, we observe a small remnant polarization, purely electronic in nature, in the absence of rotations; although this arises due to the nature of the improper mechanism, control over this property could result in `electronic-only' ferroelectrics, allowing for ultra-fast switching due to unnecessary ionic motion.

\section{Computational Methods}

All investigations were performed using density functional theory\cite{Hohenberg/Kohn:1964} as implemented in the Vienna \textit{ab-initio} Simulation Package (\textsc{VASP}).\cite{Kresse/Hafner:1993,Kresse/Hafner:1996}
We used projector augmented-wave (PAW) potentials\cite{Blochl:1994} with the 
revised generalized-gradient approximation PBE functional.\cite{PBE1:1996,PBE2:1997}
A plane-wave cutoff of 550 eV and a 2$\times$4$\times$1 Monkhorst-Pack mesh\cite{Monkhorst/Pack:1976} was used during the structural relaxations.
Following the approach of Fukushima \emph{et al.}\cite{Fukushima_etal:2011} we applied a Hubbard $U$ correction of 4 eV using the Dudarev formalism\cite{Dudarev:1998} to treat the correlated Mn $3d$ states, where we also apply antiferromagnetic collinear spin ordering.
All nine compounds were initially fixed to the experimentally determined NaLaMnWO$_6$ $P2_1$ structure, followed by a relaxation of the lattice parameters and internal atomic positions.
The electric polarization was calculated using the Berry phase method as implemented in \textsc{VASP}.\cite{King-Smith/Vanderbilt:1993,RevModPhys.66.899}
The mode decomposition was performed using the ISODISTORT tool of the ISOTROPY software suite.\cite{ISODISTORT}
Graphical rendering of the crystal structures was performed using VESTA.\cite{VESTA}

\begin{table}[b]
\centering
\caption{\label{tab:crystallographic} Equilibrium lattice parameters and monoclinic angle ($\alpha$) of the nine $AA^{\prime}$MnWO$_6$ ordered superlattices investigated. All structures were constrained to the $P2_1$ space group (\#4).
} 
\begin{tabular*}{0.5\textwidth}{@{\extracolsep{\fill}}llllll}
\hline\hline
$A$ & $A^{\prime}$ & a (\AA) & b (\AA) & c (\AA) & $\alpha$ (deg.) \\
\hline
Na & La & 5.572 & 5.597 & 8.016 & 90.225 \\
Na & Nd & 5.541 & 5.667 & 8.034 & 90.303 \\
Na & Y & 5.438 & 5.637 & 7.961 & 90.270 \\
K & La & 5.696 & 5.722 & 8.237 & 90.504 \\
K & Nd & 5.634 & 5.718 & 8.243 & 90.587 \\
K & Y & 5.519 & 5.692 & 8.254 & 90.729 \\
Rb & La & 5.737 & 5.752 & 8.337 & 90.549 \\
Rb & Nd & 5.545 & 5.635  & 9.492 & 92.227 \\
Rb & Y & 5.407 & 5.594 & 9.721 & 93.131 \\
\hline\hline
\end{tabular*}
\end{table}

\section{Results}
\subsection{Ground State Structures}
Nine double perovskites were generated by occupying 
alternating $A$- and $B$-sites along the [001] and [111] crystallographic directions with 
monovalent alkali metal ($A$) and trivalent ($A^{\prime}$) cations, and Mn and W, respectively to give chemical 
formula $AA^{\prime}$MnWO$_6$. 
We refer to each compound according to the A-site chemistry 
as $AA^{\prime}$MW, where the layered combinations investigated include: 
NaLaMW, NaNdMW, NaYMW, KLaMW, KNdMW, KYMW, RbLaMW, RbNdMW, and RbYMW.

Of these manganese tungstates, NaLaMnWO$_6$, NaNdMnWO$_6$, and KLaMnWO$_6$ have been experimentally realized and ordered in the layered and rock salt configuration.\cite{King/Woodward_etal:2007,King/Wills/Woodward:2009,GarciaMartin_etal:2011}
While NaLaMW and NaNdMW exhibit the polar $P2_1$ space group and an $a^-a^-c^+$ tilt pattern, KLaMW is found to be noncentrosymmetric-non-polar ($P\bar{4}2m$) and has a complex incommensurate tilt pattern.\cite{GarciaMartin_etal:2011}
However, in order to investigate the magnitude of the spontaneous polarization and make meaningful comparisons between members of this family, we constrain the lattice relaxations of all compounds to 
be within the $P2_1$ space group and caution synthetic researchers that this phase is unlikely to be the ground state.
In each case, the electric polarization is along the $b$ direction.

\begin{figure}[t]
\centering
\includegraphics[width=1.0\columnwidth,clip]{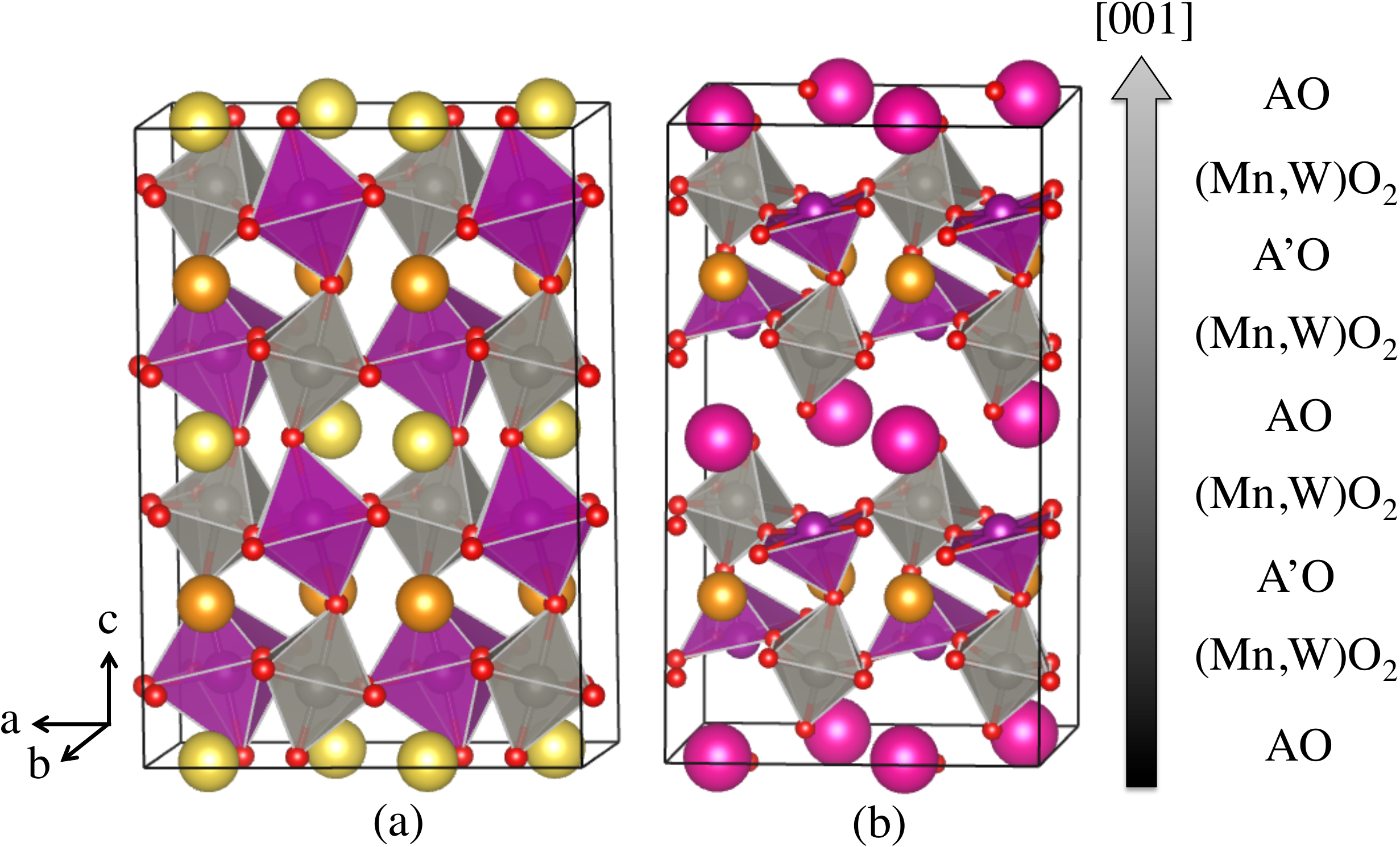}\vspace{-0.5\baselineskip}
\caption{The crystal structure of (a) NaNdMnWO$_6$ and (b) RbNdMnWO$_6$. The Na, Rb, and Nd atoms are yellow, pink, and orange, respectively. Each structure exhibits the $P2_1$ space group, and contains rock salt ordered Mn (maroon) and W (grey) $B$-sites as well as layered $A$ and $A^\prime$ sites. The individual layers of the structures along [001] are labelled as shown on the schematic to the right.
}
\label{fig:structures}
\end{figure}

The structure of a typical member of this family is shown in Figure \ref{fig:structures}a.
We find that while the lattice parameters and monoclinic angles are relatively independent of the $A^{\prime}$ chemistry, they increase with increasing $A$ cation size (Table \ref{tab:crystallographic}).
For seven of the nine compounds, the lattice parameters are relatively similar and the monoclinic angle $\alpha$ is less than 1$^{\circ}$.
Interestingly, the RbNdMW and RbYMW compounds are found to have highly elongated 
$c$-axes and large monoclinic angles (see the last two rows of Table \ref{tab:crystallographic}).
Together these cell distortions also elongate the Mn-O bonds aligned along the $c$-axis to the point that they are essentially ``broken.'' 
Figure \ref{fig:structures}b clearly shows this effect, where the Mn coordination transforms to a square pyramidal geometry.
Although one might want to attribute this effect solely to the increase in the ionic radius of Rb 
compared to the other alkali metals, RbLaMW does not exhibit this structure type.
Rather we find that the controlling factor is 
the \emph{difference} in the size of the $A$-site ionic radii ($r$) of the cations, \emph{i.e.}, $\Delta r = |r_A-r_{A^\prime}|$.
Only above a critical $\Delta r$ ($\sim 0.45$ \AA) do the $AA^{\prime}$MW compounds 
adopt the highly distorted structure.

\begin{table}
\centering
\caption{\label{tab:bvs} Bond valence (BV) sums of $A$-O, $A^{\prime}$-O, Mn-O, and W-O bonds of all nine $AA^{\prime}$MnWO$_6$ double perovskites. The values were computed using $BV = \exp[(R_\mathrm{o}-R)/{B}]$, where $R_\mathrm{o}$ is a parameter describing the bond length when atoms have their ideal valence,\cite{Brown/Altermatt:1985} $R$ is the average bond length determined from DFT calculations, and $B$ is an empirical constant (set to 0.37 for each case here).
} 
\begin{tabular*}{0.5\textwidth}{@{\extracolsep{\fill}}llllll}
\hline\hline
$A$ & $A^{\prime}$ & $A$-O & $A^{\prime}$-O & Mn-O & W-O  \\
\hline
Na & La & 1.15 & 4.39 & 2.33 & 5.38 \\
Na & Nd & 1.13 & 4.14 & 2.05 & 5.42 \\
Na & Y & 1.20 & 4.14 & 2.05 & 5.43 \\
K & La & 1.19 & 3.98 & 2.01 & 5.48 \\
K & Nd & 1.18 & 3.89 & 1.97 & 5.44 \\
K & Y & 1.14 & 3.96 & 1.91 & 5.42 \\
Rb & La & 1.49 & 3.78 & 1.92 & 5.48 \\
Rb & Nd & 0.91 & 3.88  & 1.23 & 5.30 \\
Rb & Y & 0.95 & 3.91 & 1.07 & 5.25 \\
\hline\hline
\end{tabular*}
\end{table}

An examination of the bond valence sums (Table \ref{tab:bvs}) also shows that the Mn atoms in RbNdMW and RbYMW are highly undercoordinated, decreasing from their nominal oxidation state of $2+$ to nearly $1+$.
Although this analysis suggests that these two structures are unlikely to be synthesized in this polymorph, we keep the structures in our hypothetical suite of compounds to evaluate how the electric polarization evolves with $A$ cation substitution. 
Additionally, the $A$-site bond valence in RbLaMW suggests that the Rb atom is over coordinated in this compound.
This occurs because Rb and La are the largest alkali and rare earth cations investigated, respectively, leading to RbLaMW exhibiting the largest octahedral rotations; this in turn gives a higher coordination number to Rb in comparison to the $A$-sites of the other compounds.

\subsection{Dielectric and Magnetic Properties}

\begin{table}[b]
\centering
\caption{\label{tab:bulkcompounds}The ionic radii of 12-fold coordinated $A$ ($r_A$) and $A^{\prime}$ ($r_{A^{\prime}}$) atoms, polarization ($P$), and energy difference between the $P2_1$ ground state and the lowest energy centrosymmetric, supergroup $P2_1/m$, phase ($E_B$) for all nine $AA^{\prime}$MnWO$_6$ compounds.
} 
\begin{tabular*}{0.5\textwidth}{@{\extracolsep{\fill}}llcccc}
\hline\hline
$A$ & $A^{\prime}$ & $r_A$ (\AA) & $r_{A^{\prime}}$ (\AA) & $P$ ($\mu$C/cm$^2$) & $E_B$ (meV/f.u.) \\
\hline
Na & La & 1.39 & 1.36 & 16.1 & 46.0 \\
Na & Nd & 1.39 & 1.27 & 19.3 & 224 \\
Na & Y & 1.39 & 1.25 & 23.6 & 426 \\
K & La & 1.64 & 1.36 & 15.5 & 37.8 \\
K & Nd & 1.64 & 1.27 & 20.8 & 109 \\
K & Y & 1.64 & 1.25 & 26.1 & 353 \\
Rb & La & 1.72 & 1.36 & 15.1 & 20.6 \\
Rb & Nd & 1.72 & 1.27  & 19.0 & 193 \\
Rb & Y & 1.72 & 1.25 & 23.6 & 243 \\
\hline\hline
\end{tabular*}
\end{table}
 
\begin{figure}[t]
\centering
\includegraphics[width=1.0\columnwidth,clip]{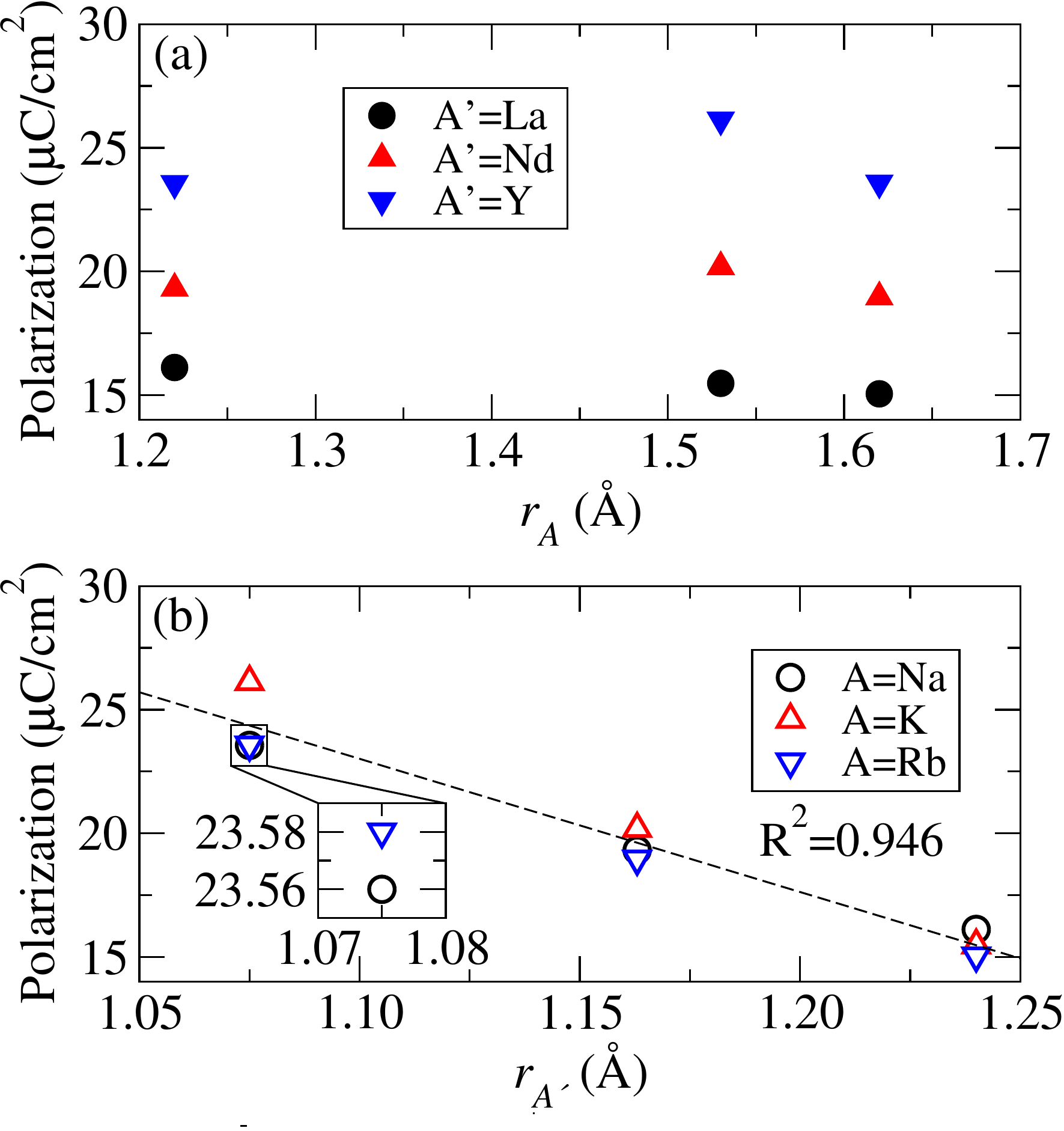}\vspace{-0.5\baselineskip}
\caption{The total polarization of each $AA^{\prime}$MnWO$_6$ compound as a function of the size of the (a) alkali $A$ cation and (b) rare earth $A^{\prime}$ cation. While the polarization remains relatively constant upon substitution of the $A$ site, it is strongly correlated to the size of the rare earth cation. The dashed line shows a linear fit to the data, while R$^2$ is the coefficient of determination for the fit.
}
\label{fig:polarization}
\end{figure}

Table \ref{tab:bulkcompounds} reports the computed electric polarizations for the ground state 
structures of all nine compounds.
The polarization does not show significant variation when changing the alkali cation and keeping the rare-earth chemistry fixed (Figure \ref{fig:polarization}a), but it markedly increases upon substitution of the trivalent $A^\prime$ cation with a smaller lanthanide element at a fixed alkali metal composition (Figure \ref{fig:polarization}b).
This suggests that the size of the rare earth cation (Table \ref{tab:bulkcompounds}) influences the magnitude of the electric polarization to a greater degree than the alkali metal.
However, this does not mean that the $A$-site does not influence the magnitude of the polarization; there are several subtle effects which are not explained by the rare earth chemistry of the $A^{\prime}$-site alone and will be explored in more detail in the next section. 
Additionally, there is a clear trend between the polarization and the energy difference between the $P2_1$ ferroelectric and $P2_1/m$ paraelectric phases (given as $E_B$ in Table \ref{tab:bulkcompounds}), which can be used as an estimation of the intrinsic ferroelectric switching barrier.

\begin{figure}[h!]
\centering
\includegraphics[width=0.9\columnwidth,clip]{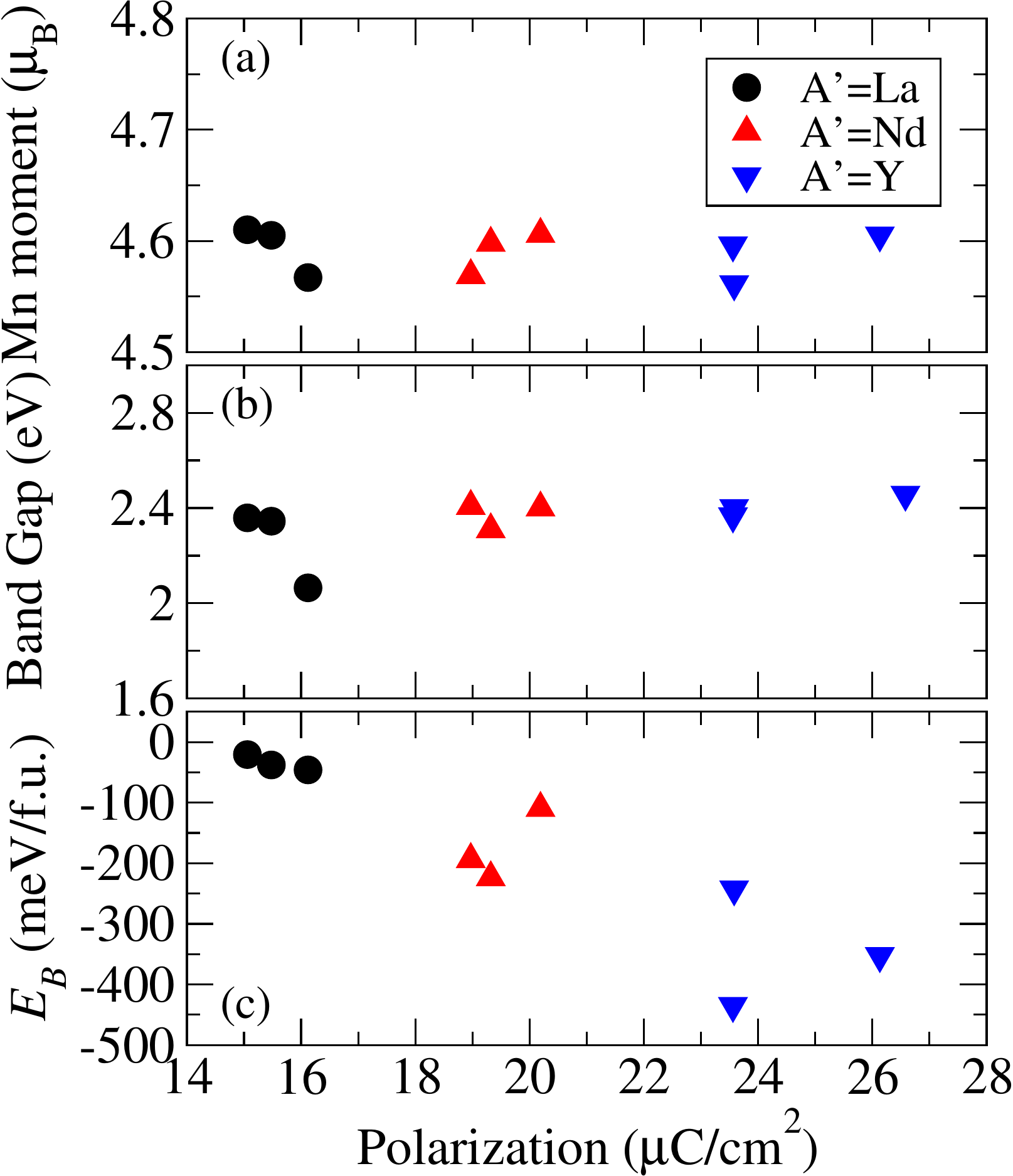}\vspace{-0.5\baselineskip}
\caption{The (a) Mn magnetic moment, (b) band gap, and (c) energy difference between the $P2_1$ and $P2_1/m$ phases of each compound. While the magnetic moment and band gap remain relatively unchanged upon chemical substitution, the energy difference between the phases increases substantially as the polarization increases. 
}
\label{fig:props}
\end{figure}

We next investigate how the Mn magnetic moment and band gap evolves with respect to 
chemical substitution on the $A$-site.
We find there is no significant correlation in either of these quantities as a function of polarization.
Figure \ref{fig:props}a and \ref{fig:props}b show that there is small variation between the magnetic properties and electronic band gap with  $A$-site chemistry---both properties are largely unaffected by the alkali metal and trivalent cations on the $A$-site.
However, the properties are closely clustered for a given $A^{\prime}$ cation.
We understand this behavior as follows: the magnetic moment only depends on the Mn atoms, and any small changes to it are induced by inductive effects from the $A$ cations that weakly change the length and covalency of the Mn--O bond.\cite{Cammarata/Rondinelli:2014}
The electronic density of states (not shown) also primarily consists of O $2p$-states at the top of the valence band and W $5d$-states at the bottom of the conduction band. 
The $A$ and $A^\prime$ cations contribute states far from these band edges, and as a result their chemistry does not significantly impact the band gap.

In order to switch the polarization in these rotationally-induced improper ferroelectrics, the entire sense of the in-phase tilts must be reversed. 
The $P2_1/m$ structure is the mostly likely intermediate centric phase along the switching path as it contains no in-phase rotations.\cite{Fukushima_etal:2011}
As the size of the rare earth cation decreases (and thus the polarization increases), this energy difference between these two phases increases by a large amount; this is also reflected in the fact that the compounds with smaller cations have larger octahedral rotations, as mentioned previously.
Thus, while an enhancement in the polarization can be achieved through $A$-site chemical substitution, there is a trade off in the ability of the ferroelectric polarization to be switched.
This was reported previously 
in other rotation-driven ferroelectrics such as layered perovskites and Ruddlesden-Popper phases.\cite{Mulder/Rondinelli/Fennie:2013}

\section{Discussion}
\subsection{Crystal-Chemistry Descriptors for Electric Polarization}

To build a more thorough microscopic understanding of the evolution in the electric polarization with $A$ and $A^{\prime}$ cation substitution, we explore how structural-chemistry descriptors relate the dielectric properties to the crystal structure.
One of the most common descriptors for perovskite oxides is the Goldschmidt tolerance factor,\cite{Goldschmidt:1926} which for the double perovskites examined here can be expressed as 
\begin{equation}
\tau_\mathrm{avg} = \frac{\bar{r}_A+r_\mathrm{O}}{\sqrt{2}\,(\bar{r}_B+r_\mathrm{O})},
\end{equation}
where $\bar{r}_{A}$ and $\bar{r}_{B}$  are the average Shannon ionic radii of the $A$-site and $B$-site cations of 12-fold coordinated $A$ and $A^{\prime}$ and 6-fold coordinated $B$ and $B^{\prime}$ atoms.\cite{Shannon:1976}
This quantity gives a simple measure of the distortion of a perovskite oxide; the closer $\tau_\mathrm{avg}$ is to 1, the less distorted (or more cubic) the structure tends to be.
If $\tau_\mathrm{avg}$ is less than 1, then the $A$-site cations are too small for the interstices between $B$O$_6$ octahedra, and the extended octahedral network rotates in order to alleviate this underbonding of the $A$-site species.

\begin{figure}[]
\centering
\includegraphics[width=1.0\columnwidth,clip]{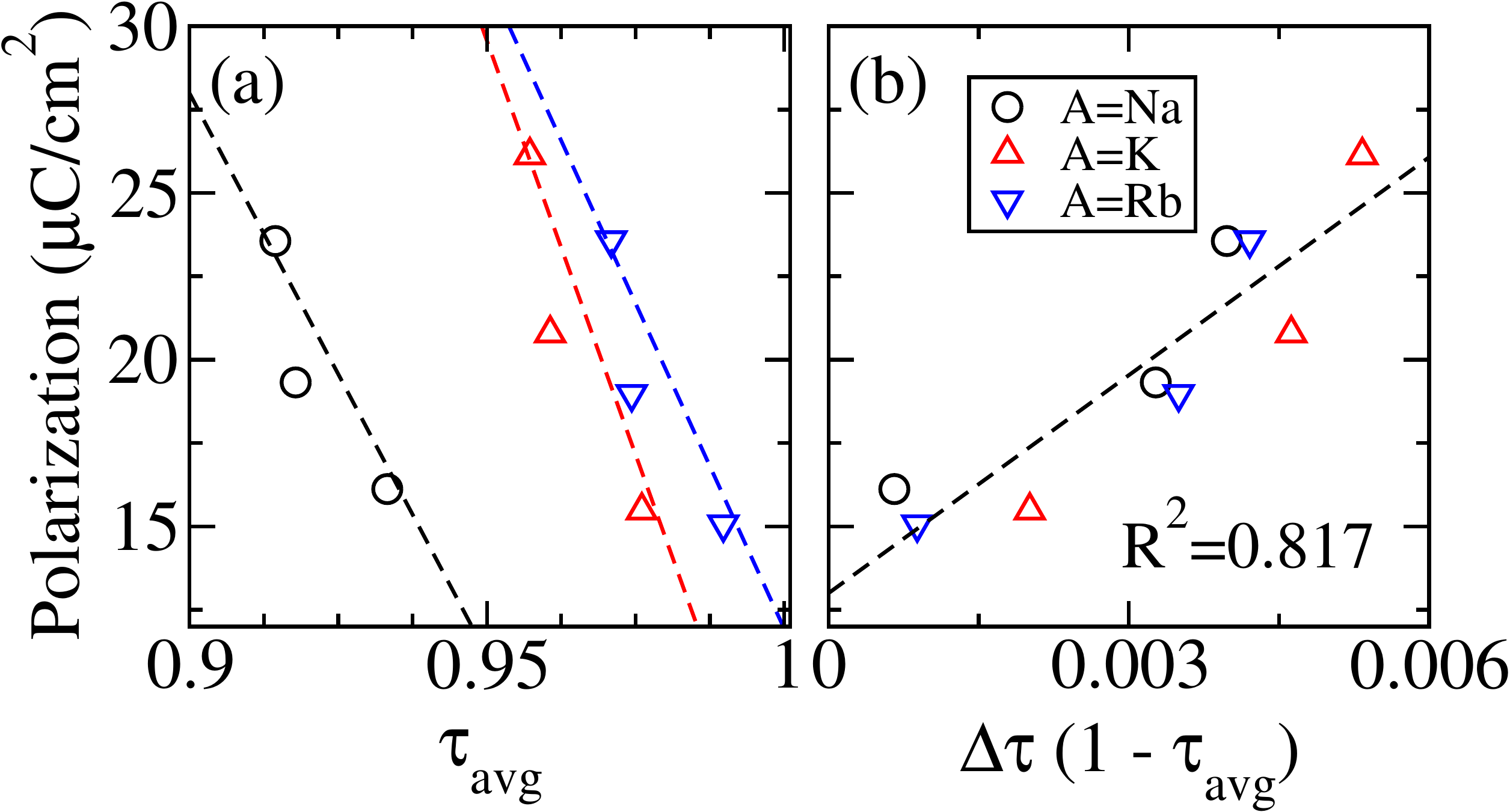}\vspace{-0.7\baselineskip}
\caption{(a) Polarization versus tolerance factor of the nine $AA^{\prime}$MnWO$_6$ compounds, computed using the Shannon ionic radii of 12-fold coordinated $A$ and $A^{\prime}$ sites. While the tolerance factor by itself is not a good overall predictor of the polarization, it is a good descriptor within each family of constant rare earth $A^{\prime}$ site (dashed lines are linear fits to the data). (b) Polarization versus tolerance factor renormalized to the tendency of $A$-sites to displace (see text). This provides a much better prediction of the polarization among all compounds.
}
\label{fig:tolerance}
\end{figure}

Because of the ubiquity of this quantity in characterizing perovskites, we first examined the effect of $\tau_\mathrm{avg}$ on the polarization.
This parameter alone does not well describe the evolution in the magnitude of the polarization for all compounds and is anti-correlated with the polarization within a given trivalent $A^\prime$ family (Figure \ref{fig:tolerance}a).
Previous work has identified that the difference in tolerance factor ($\Delta \tau$) of the bulk $AB$O$_3$ constituents is also an important descriptor of the polarization in perovskites with layered $A$ cation order exhibiting hybrid improper ferroelectricity.\cite{Mulder/Rondinelli/Fennie:2013}
Starting from a symmetry analysis, Mulder \textit{et al.} derived a relationship between spontaneous polarization and tolerance factor of the form $P \approx \Delta \tau (1-\tau_\mathrm{avg})$.
Here, we consider $\Delta \tau$ to be the difference between the average tolerance factor of the $(A,A^{\prime})$MnO$_3$ and $(A,A^{\prime})$WO$_3$ bulk compounds.
This quantity improves the correlation among all compounds and the spontaneous polarization (Figure \ref{fig:tolerance}b) due to information about the difference in atomic radii being included, which will be shown to be an important factor in determining the magnitude of the polarization.
Nonetheless, the ionic size of the $A^{\prime}$ cation alone provides a better predictor of the polarization (Figure~\ref{fig:polarization}b), albeit is anti-correlated.

\begin{table}[t]
\centering
\caption{\label{tab:modedecomp}Amplitude of the three primary modes relating the nine $P2_1$ $AA^{\prime}$MnWO$_6$ ordered superlattices to the undistorted high symmetry $P4/nmm$ phase. Labels of the irreducible representations are generated with respect to the $P4/nmm$ structure.
} 
\begin{tabular*}{0.5\textwidth}{@{\extracolsep{\fill}}llccc}
\hline\hline\\[-0.95em]
$A$ & $A^{\prime}$ & $Q_{\Gamma_5^+}$ (\AA) & $Q_{\Gamma_1^-}$ (\AA) & $Q_{\Gamma_5^-}$ (\AA) \\[0.25em]
\hline
Na & La & 1.336 & 0.897 & 0.668 \\
Na & Nd & 1.486 & 1.986 & 0.899 \\
Na & Y & 1.589 & 1.273 & 1.088 \\
K & La & 0.984 & 0.658 & 0.449 \\
K & Nd & 1.079 & 0.861 & 0.625 \\
K & Y & 1.251 & 1.131 & 0.829 \\
Rb & La & 0.868 & 0.499 & 0.358 \\
Rb & Nd & 1.236 & 0.882 & 0.499 \\
Rb & Y & 1.503 & 1.109 & 0.714 \\
\hline\hline
\end{tabular*}
\end{table}

\begin{figure}[h!]
\centering
\includegraphics[width=0.9\columnwidth,clip]{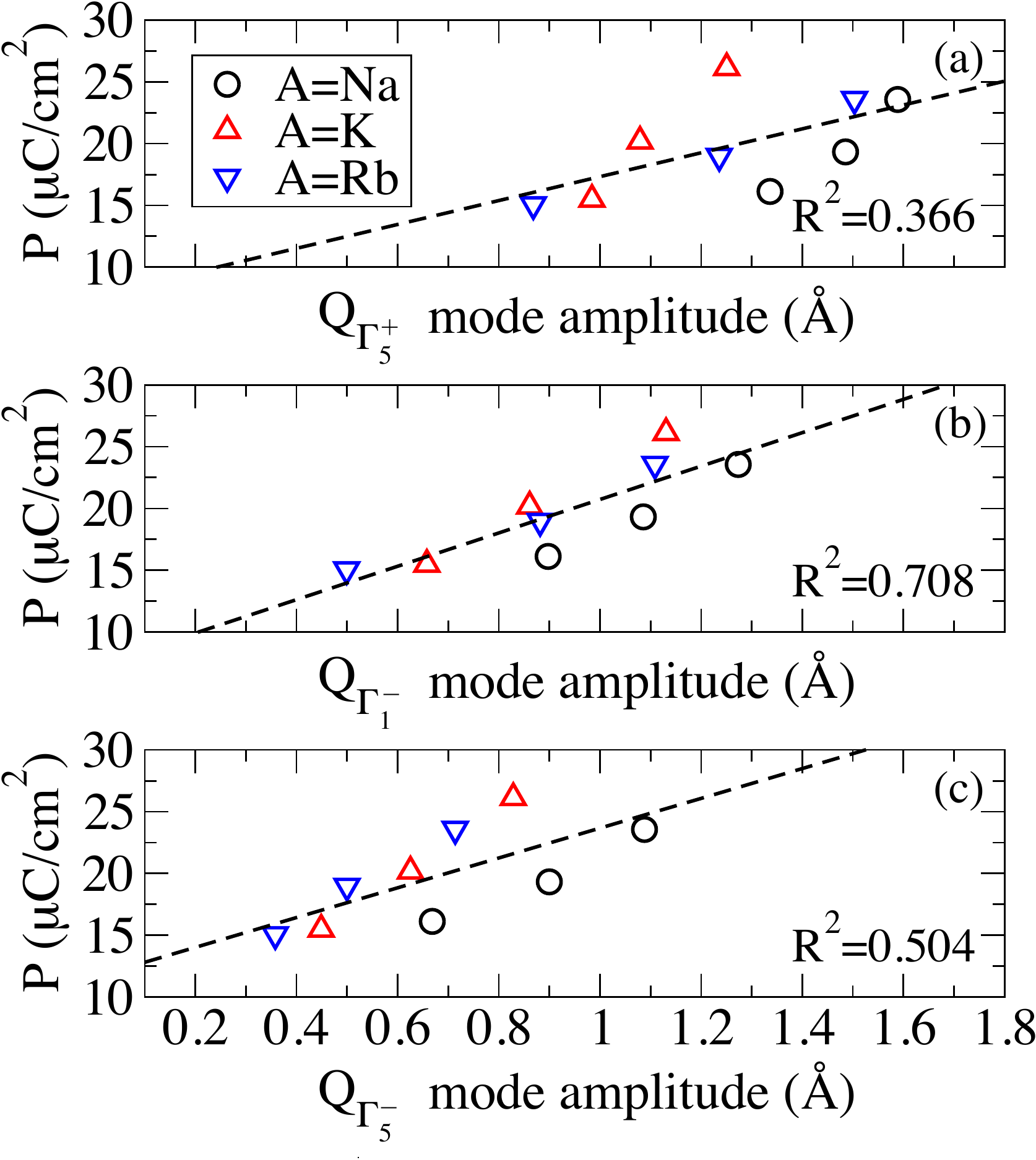}\vspace{-1\baselineskip}
\caption{Polarization as a function of (a) $\Gamma_5^+$ (out-of-phase rotations), (b) $\Gamma_1^-$ (in-phase rotations), and (c) $\Gamma_5^-$ (polar mode) mode amplitudes of each $AA^{\prime}$MW compound.
}
\label{fig:modesvsp}
\end{figure}

We next perform a mode decomposition of the polar structures.
Three primary modes relating the low symmetry ground state $P2_1$ structure to the undistorted high symmetry $P4/nmm$ structure have been previously identified and are described by the following irreducible representations (irreps): $\Gamma_5^+$ (describing the out-of-phase rotations), $\Gamma^-_1$ (describing the in-phase rotations), and $\Gamma_5^-$ (polar mode).\cite{Fukushima_etal:2011}
(Note the mechanical representations of $\Gamma_1^-$ and $\Gamma^+_5$ are similar to irreps $M_3^+$ and $R_4^+$ irreps found in the perovskite literature, which are responsible for reducing a cubic $Pm\bar{3}m$ $AB$O$_3$ perovskite to the common $Pnma$ orthorhombic phase).
The mode amplitude $Q$ provides a measure of the degree to which a structure is distorted by these three modes (Table \ref{tab:modedecomp}).
We find that the amplitude of $\Gamma_5^+$ and $\Gamma^-_1$ (and thus the $B$O$_6$ and  $B^{\prime}$O$_6$ octahedral rotations) increases as a function of decreasing atomic size.
All three of the main modes show a somewhat clear trend with the polarization, the amplitude of the $\Gamma^-_1$ mode is the most strongly correlated (Figure 6). This behavior suggests that the out-of-phase rotations, exhibit the largest control over the polarization.
Additionally, the polar $\Gamma_5^-$ mode follows the same trend, indicating that the magnitude of the spontaneous polarization should increase as the $A^{\prime}$ atomic species get smaller (Fig.~\ref{fig:modesvsp}).
This is to be expected given the hybrid improper nature of the polarization, as the net dipole in these materials arises due to a non-cancellation of $A$-site cation displacements, \textit{i.e.}, having smaller atoms means they can undergo larger displacements.

By taking all of this structural information into consideration, KYMnWO$_6$ appears to be the chemistry best suited for polarization enhancement and experimental investigation in this family of materials.
The difference in the atomic size of the K and Y ($\Delta r$) is large enough to maximize the polarization, while also small enough to maintain the desired perovskite structure (\textit{i.e.} the bond valence of the Mn-O remains close to the nominal value of 2+, Table \ref{tab:bvs}).
As mentioned previously, however the size of the rotations also increases the ferroelectric switching barrier, due to the fact that reversal of the in-phase rotations is required for switching of the polarization ($E_B$, Table \ref{tab:bulkcompounds}).

\begin{figure}[h!]
\centering
\includegraphics[width=1.0\columnwidth,clip]{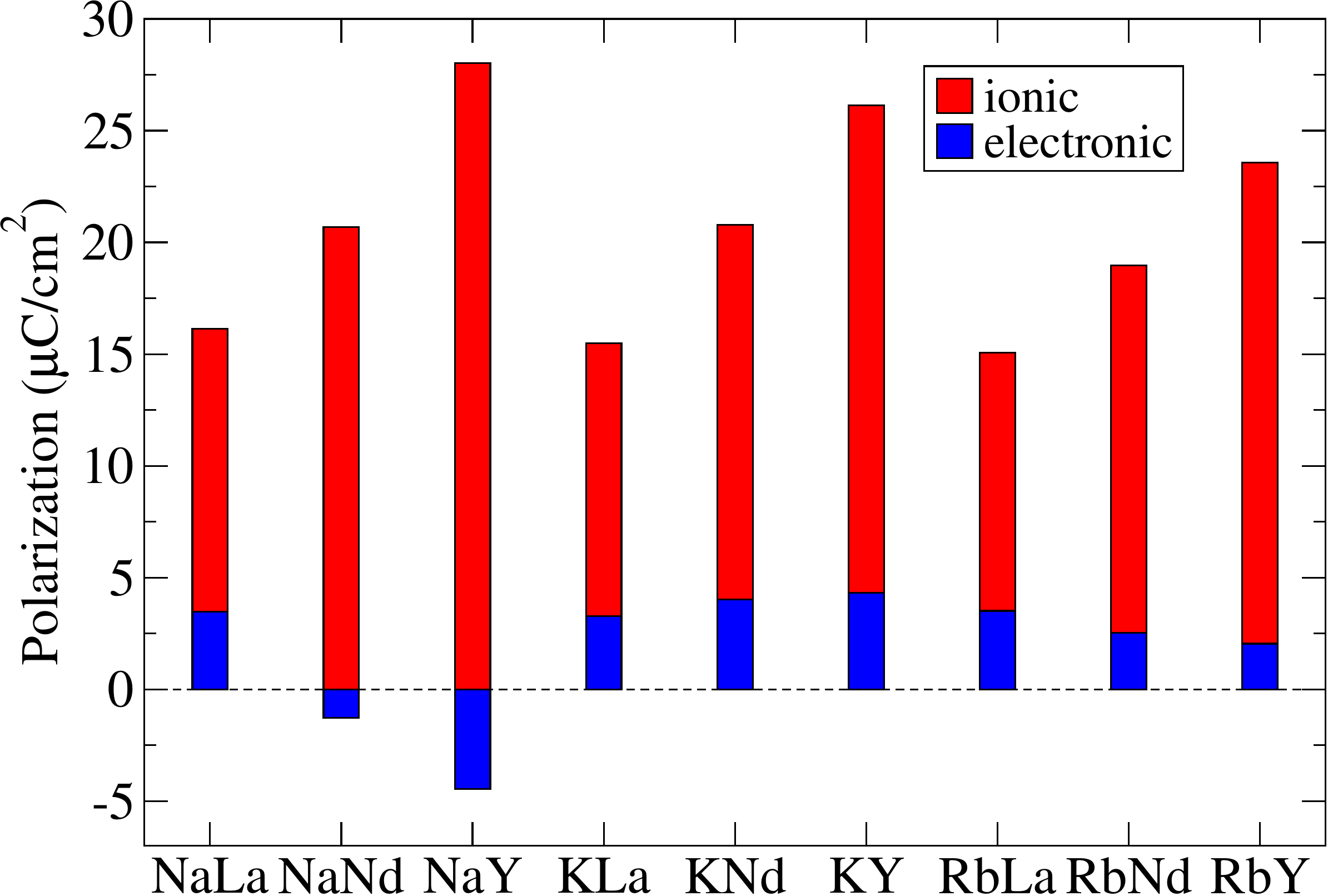}\vspace{-0.5\baselineskip}
\caption{The ionic (red) and electronic (blue) contributions to the total electric polarization for each of the nine double perovskite compounds.
}
\label{fig:Pdecomp}
\end{figure}

\subsection{Contributions to the Electric Polarization}

The total spontaneous polarization can be decomposed into two parts: an ionic contribution from atomic displacements ($P_{ion}$), and an electronic contribution arising from displacement of the electron cloud ($P_{elec}$), as shown in Figure \ref{fig:Pdecomp}.
While the size of the rare earth $A^{\prime}$ cation largely controls the magnitude of the spontaneous polarization, an investigation of the contributions of each atomic species can lead to further insights.
Note that while this division is useful for a qualitative discussion and analysis of the polarization, only the sum of the two (\textit{i.e.} the total polarization $P=P_{ion}+P_{elec}$) has any real physical meaning.

We first examine the ionic contribution  (Figure \ref{fig:Pdecomp}, red bars).
In these hybrid improper ferroelectrics we would anticipate a smaller $A^{\prime}$ cation to result in a larger polarization due to its ability to displace further.
Additionally, a larger alkali atom should also result in a larger polarization, because the $A$ cation displaces less in the direction opposite to the $A^{\prime}$ cation, \textit{i.e.} cancelling less of the $A^{\prime}$ polarization.
One would predict two clear trends by applying these concepts to the manganese tungstates \textit{a priori}: (i)  $P_{A\mathrm{LaMW}} < P_{A\mathrm{NdMW}} < P_{A\mathrm{YMW}}$ and 
(ii) $P_{\mathrm{Na}A^{\prime}\mathrm{MW}} < P_{\mathrm{K}A^{\prime}\mathrm{MW}} < P_{\mathrm{Rb}A^{\prime}\mathrm{MW}}$.
NaLaMnWO$_6$ would therefore have the smallest polarization while RbYMnWO$_6$ would have the largest, with the other seven compounds found between these extremes.
While this expectation holds for the rare earth trend (Figure \ref{fig:polarization}a), it does not hold for the alkali metal trend (Figure \ref{fig:polarization}b).
If $A^{\prime}$=La, we find that $P_{\mathrm{RbLaMW}} < P_{\mathrm{KLaMW}} < P_{\mathrm{NaLaMW}}$, but when La is replaced with Nd, the trend changes such that $P_{\mathrm{RbNdMW}} < P_{\mathrm{NaNdMW}} < P_{\mathrm{KNdMW}}$.

\begin{figure*}[t]
\centering
\includegraphics[width=1.45\columnwidth,clip]{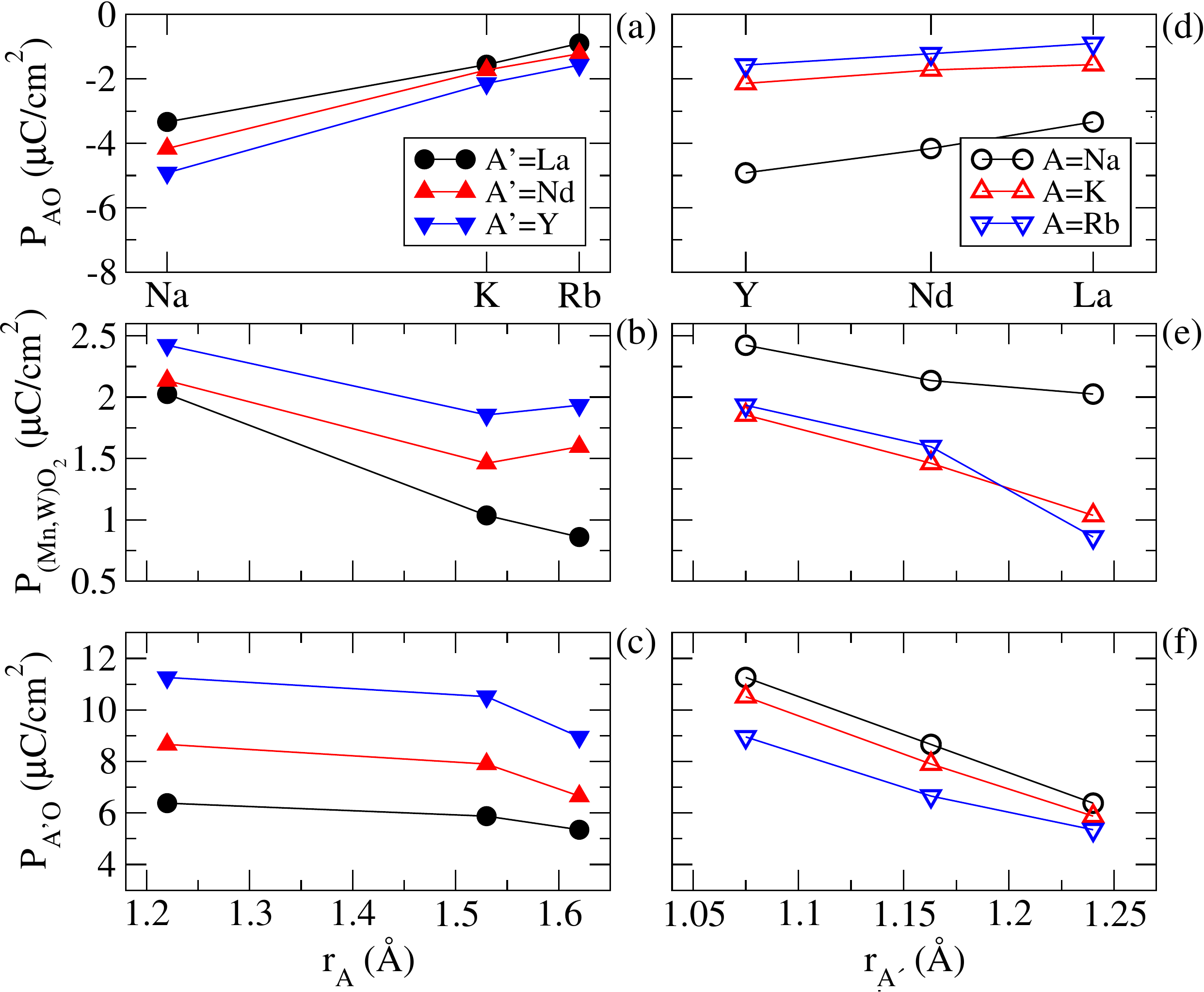}\vspace{-1\baselineskip}
\caption{Evolution of the ionic contribution to the polarization of the $A$O, $B$O$_2$, and $A^{\prime}$O layers of each $AA^{\prime}$MnWO$_6$ compound as a function of the $A$-site ($r_A$, a--c) and $A^{\prime}$-site ($r_{A^{\prime}}$, d--f) radius. See text for a detailed discussion.
}
\label{fig:decomp}
\end{figure*}

We next investigate these discrepancies by decomposing the polarization into contributions from each layer (Figure \ref{fig:decomp}).
As can be seen in Figure \ref{fig:structures}, each compound consists of two $A$O layers, two (Mn,W)O$_2$ layers, and two $A^{\prime}$O layers along [001].
The polarization of one of each of these layers in the nine superlattices is plotted as a function of $A$ cation size (Figure \ref{fig:decomp}a-c) or $A^{\prime}$ cation size (Figure \ref{fig:decomp}d-f).
Figure~\ref{fig:decomp}a reveals the trend that larger $A$-site cations result in a lower $A$O layer polarization (and therefore a smaller cancellation of the $A^{\prime}$ layer) occurs as expected.
Additionally, the (Mn,W)O$_2$ layer polarization decreases as a function of increasing $A$ cation size (Figure \ref{fig:decomp}b) due to the smaller $A$-site allowing for greater $B$-site displacements; the increase in the 
layer polarization for the RbNd and RbY compounds occurs from the Mn--O bond breaking.
Also as expected, there is a weak dependence of the rare earth layer polarization
on the alkali metal on the size  (Figure \ref{fig:decomp}c), and \emph{vice versa} (Figure \ref{fig:decomp}d).
Interestingly, the contribution of the (Mn,W)O$_2$ layer to the total polarization switches to become greater in the $A$=K than in the $A$=Rb compounds upon substitution of Nd by La (Figure \ref{fig:decomp}e). 
By looking at the atomically resolved (Mn,W)O$_2$ layers, we find that the contributions of the Mn and W atoms to the polarization is larger in the case of $A$=K than $A$=Rb.
The contributions of the oxygen atoms are larger in KLaMW than RbLaMW because the octahedral rotations are also larger; in RbNdMW and RbYMW, however, the octahedral rotations become larger than KNdMW and KYMW due to the separation of the layers, thus allowing for a larger contribution of its oxygen atoms.
As expected, the layer polarization of the $A^{\prime}$O layer increases as a function of decreasing atomic size (Figure \ref{fig:decomp}f).

Following this analysis, we can now explain the trends in the polarization seen in Figure \ref{fig:polarization}b.
In the $A^{\prime}$=La family, the smaller alkali cation should result in a larger cancellation of the LaO layer polarization, which remains nominally the same in each of the three compounds.
The fact that the (Mn,W)O$_2$ layers contribute more to the total polarization with smaller $A$-sites overcomes this cancellation, resulting in the seen trend of $P_{\mathrm{NaLaMW}} > P_{\mathrm{KLaMW}} > P_{\mathrm{RbLaMW}}$.
When La is substituted with Nd, the amount that $P_{\mathrm{Na}}$ cancels $P_{\mathrm{Nd}}$ is much greater than the amount $P_{\mathrm{K}}$ cancels $P_{\mathrm{Nd}}$, even more so than when $A^{\prime}$=La.
This, in combination with the trend of the (Mn,W)O$_2$ layer polarization gives $P_{\mathrm{KNdMW}} > P_{\mathrm{NaNdMW}} > P_{\mathrm{RbNdMW}}$.
Finally, this same reasoning can rationalize the polarization trend seen in the $A^{\prime}$=Y family.

While the bulk of the total polarization comes from atomic displacements as discussed above ($P_{ion}$), the \emph{electronic} component provides a non-negligible contribution.
In seven of the nine structures, the two contributions are of the same sign, with the electronic part providing an enhancement of 2 to 5 $\mu$C/cm$^2$.
In the NaNd and NaY compounds, however, the electronic polarization is opposite to that of the ionic contribution, resulting in a decrease of the total polarization.
Interestingly, these two materials also have the smallest average $A$-site radius.
However, because the magnitude of the electronic polarization in the other compounds are relatively similar, this does not provide a complete explanation and requires further investigation.

\begin{figure}[t]
\centering
\includegraphics[width=1.0\columnwidth,clip]{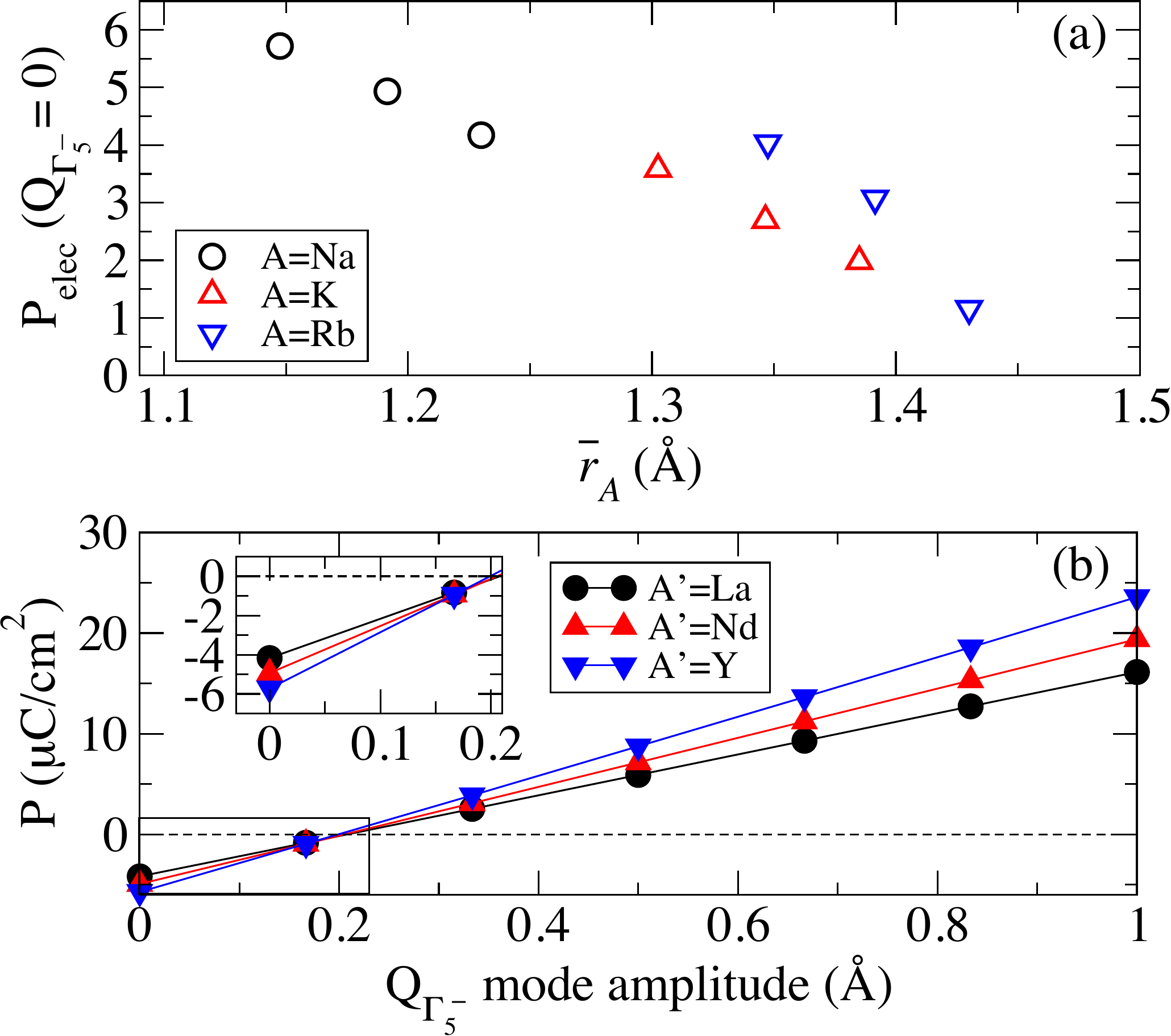}\vspace{-0.7\baselineskip}
\caption{(a) The spontaneous electronic polarization remaining when Q$_{\Gamma_5^-}=0$ (no polar ionic displacements) in all nine $AA^{\prime}$MnWO$_6$ compounds as a function of average $A$-site atomic radius. (b) Evolution of the total polarization as a function of Q$_{\Gamma_5^-}$ in the family of Na$A^{\prime}$MnWO$_6$ compounds. At Q$_{\Gamma_5^-}$=0, the polarization is purely electronic in nature.
}
\label{fig:electronic_decomp}
\end{figure}

An electronic contribution also is present even if the polar displacements are removed from the crystal structures (Fig.~\ref{fig:electronic_decomp}).
Because the $\Gamma_1^-$ and $\Gamma_5^+$ modes (in-phase and out-of-phase rotations, respectively) alone are sufficient to lift inversion in the presence of layered $A$-sites and rock salt ordered $B$-sites (that is, the space group remains $P2_1$ without the $\Gamma_5^-$ mode), these materials could theoretically still exhibit a spontaneous polarization without the ionic displacements.
To investigate this aspect, we decreased the $Q_{\Gamma_5^-}$ amplitude while keeping the magnitude of $Q_{\Gamma_5^+}$ and $Q_{\Gamma_1^-}$ fixed to the values found in the equilibrium ground state structure.
When $Q_{\Gamma_5^-}=0$, an electronic polarization ($P_{elec}$) remained (Figure \ref{fig:electronic_decomp}a) which was aligned opposite  to the total polarization found in the $Q_{\Gamma_5^-}=1$ ground state.
Additionally, as $Q_{\Gamma_5^-}$ increases, the total polarization evolves smoothly to the value obtained in the ground state structures (Figure \ref{fig:electronic_decomp}b).
This result seems counter to what the trilinear coupling term in the free energy shown previously implies, which is that the total polarization should go to zero if the polar mode goes to zero; however, the simplification of the phenomenological invariant disguises the fact that there are both ionic and electronic contributions to this invariant (such a distinction could also allow for a new interpretation of electronically-induced improper ferroelectrics, such as HoMnO$_3$ in terms of generalized spin rotations).
This is shown by removing either of the rotational modes (while keeping the polar mode fixed to zero), which destroys this remnant polarization.

A large number of perovskites are cubic with no rotations and disordered $A$-site cations at high temperatures. 
If the octahedral rotations can then be maintained at some temperature above which the $A$-site cations order in such a way that the $\Gamma_5^-$ mode causes the layers to cancel the ionic contribution to the polarization, we conjecture the electronic only contribution could be observed experimentally.
We believe that this interesting effect deserves further study, as materials displaying purely electronic polarizations could result in entirely new families of ferroelectrics which undergo much less fatigue due to the lack of ionic motion during switching.

\section{Conclusion}

In this work, we investigated a series of nine $AA^{\prime}$MnWO$_6$ double perovskite oxides containing both chemically ordered $A$- and $B$-sites.
Through substitution of alkali earth ($A$=Na, K, Rb) and rare earth ($A^{\prime}$=La, Nd, Y) cations, we were able to tune the polarization from 16 to 26 $\mu$C/cm$^2$ while retaining the G-type antiferromagnetic ordering on the Mn atoms.
We then showed through a detailed structural analysis that while the tolerance factor alone is a relatively poor indicator of the size of the polarization in these materials, a better estimate can be obtained through knowledge of how the rare earth cations displace.
Further investigation of the spontaneous polarization showed that when the polar mode is removed from these structures, a remnant polarization is still present which is entirely electronic in nature.
Finally, we showed that while proper chemical substitution can increase the polarization, the energy difference between the polar $P2_1$ ground state and a high symmetry $P2_1/m$ structure (a likely switching path) also increases.
Experimental investigation, however, is needed to completely characterize the ferroelectric switching behavior.
Lastly, we note that the presence of a trilinear coupling term in the free energy often implies the presence of a magnetoelectric coupling; further investigation of how the strength of the magnetoelectric coupling and any induced weak ferromagnetism that may result varies from chemical substitution should be explored as symmetry arguments indicate that such an effect should be observed in these compounds.\cite{Fukushima_etal:2011} 
We hope that the potential for enhanced polarization shown in this work will stimulate experimental growth of these multiferroic materials.

\section{Acknowledgements}

J.Y.\ and J.M.R.\ acknowledge support from the Penn State Center for Nanoscience, DMR-0820404. 
J.Y. thanks members of the Rondinelli group for useful discussions. 
DFT calculations were performed using the CARBON cluster at the Center for Nanoscale Materials  
[Argonne National Laboratory, supported by the U.S.\ DOE, Office of Basic Energy Sciences (BES), DE-AC02-06CH11357], and the 
Extreme Science and Engineering Discovery Environment (XSEDE), which is supported by National Science Foundation grant number OCI-1053575.

\footnotesize{
\bibliography{rsc} 

\providecommand*{\mcitethebibliography}{\thebibliography}
\csname @ifundefined\endcsname{endmcitethebibliography}
{\let\endmcitethebibliography\endthebibliography}{}
\begin{mcitethebibliography}{62}
\providecommand*{\natexlab}[1]{#1}
\providecommand*{\mciteSetBstSublistMode}[1]{}
\providecommand*{\mciteSetBstMaxWidthForm}[2]{}
\providecommand*{\mciteBstWouldAddEndPuncttrue}
  {\def\EndOfBibitem{\unskip.}}
\providecommand*{\mciteBstWouldAddEndPunctfalse}
  {\let\EndOfBibitem\relax}
\providecommand*{\mciteSetBstMidEndSepPunct}[3]{}
\providecommand*{\mciteSetBstSublistLabelBeginEnd}[3]{}
\providecommand*{\EndOfBibitem}{}
\mciteSetBstSublistMode{f}
\mciteSetBstMaxWidthForm{subitem}
{(\emph{\alph{mcitesubitemcount}})}
\mciteSetBstSublistLabelBeginEnd{\mcitemaxwidthsubitemform\space}
{\relax}{\relax}

\bibitem[Eerenstein \emph{et~al.}(2006)Eerenstein, Mathur, and
  Scott]{Eerentstein/Mathur/Scott:2006}
W.~Eerenstein, N.~D. Mathur and J.~F. Scott, \emph{Nature}, 2006, \textbf{442},
  759--765\relax
\mciteBstWouldAddEndPuncttrue
\mciteSetBstMidEndSepPunct{\mcitedefaultmidpunct}
{\mcitedefaultendpunct}{\mcitedefaultseppunct}\relax
\EndOfBibitem
\bibitem[Gajek \emph{et~al.}(2007)Gajek, Bibes, Fusil, Bouzehouane,
  Fontcuberta, Barth\'{e}l\'{e}my, and Fert]{Gajek_etal:2007}
M.~Gajek, M.~Bibes, S.~Fusil, K.~Bouzehouane, J.~Fontcuberta,
  A.~Barth\'{e}l\'{e}my and A.~Fert, \emph{Nature Mater.}, 2007, \textbf{6},
  296--302\relax
\mciteBstWouldAddEndPuncttrue
\mciteSetBstMidEndSepPunct{\mcitedefaultmidpunct}
{\mcitedefaultendpunct}{\mcitedefaultseppunct}\relax
\EndOfBibitem
\bibitem[Nan \emph{et~al.}(2008)Nan, Bichurin, Dong, Viehland, and
  Srinivasan]{Nan_etal:2008}
C.-W. Nan, M.~I. Bichurin, S.~Dong, D.~Viehland and G.~Srinivasan, \emph{J.
  Appl. Phys.}, 2008, \textbf{103}, 031101\relax
\mciteBstWouldAddEndPuncttrue
\mciteSetBstMidEndSepPunct{\mcitedefaultmidpunct}
{\mcitedefaultendpunct}{\mcitedefaultseppunct}\relax
\EndOfBibitem
\bibitem[Hill(2000)]{Hill:2000}
N.~Hill, \emph{Journal of Physical Chemistry B}, 2000, \textbf{104},
  6694--6709\relax
\mciteBstWouldAddEndPuncttrue
\mciteSetBstMidEndSepPunct{\mcitedefaultmidpunct}
{\mcitedefaultendpunct}{\mcitedefaultseppunct}\relax
\EndOfBibitem
\bibitem[Varga \emph{et~al.}(2009)Varga, Kumar, Vlahos, Denev, Park, Hong,
  Sanehira, Wang, Fennie, Streiffer, Ke, Schiffer, Gopalan, and
  Mitchell]{Varga_Fennie_etal:2009}
T.~Varga, A.~Kumar, E.~Vlahos, S.~Denev, M.~Park, S.~Hong, T.~Sanehira,
  Y.~Wang, C.~J. Fennie, S.~K. Streiffer, X.~Ke, P.~Schiffer, V.~Gopalan and
  J.~F. Mitchell, \emph{Phys. Rev. Lett.}, 2009, \textbf{103}, 047601\relax
\mciteBstWouldAddEndPuncttrue
\mciteSetBstMidEndSepPunct{\mcitedefaultmidpunct}
{\mcitedefaultendpunct}{\mcitedefaultseppunct}\relax
\EndOfBibitem
\bibitem[Ghosh \emph{et~al.}(2014)Ghosh, Dey, Chakraborty, Majumdar, and
  Giri]{Ghosh_etal:2014}
A.~Ghosh, K.~Dey, M.~Chakraborty, S.~Majumdar and S.~Giri, \emph{Europhys.
  Lett.}, 2014, \textbf{107}, 47012\relax
\mciteBstWouldAddEndPuncttrue
\mciteSetBstMidEndSepPunct{\mcitedefaultmidpunct}
{\mcitedefaultendpunct}{\mcitedefaultseppunct}\relax
\EndOfBibitem
\bibitem[Li \emph{et~al.}(2014)Li, Stephens, Retuerto, Sarkar, Grams,
  Hemberger, Croft, Walker, and Greenblatt]{Li_Stephens_etal:2014}
M.-R. Li, P.~W. Stephens, M.~Retuerto, T.~Sarkar, C.~P. Grams, J.~Hemberger,
  M.~C. Croft, D.~Walker and M.~Greenblatt, \emph{Journal of the American
  Chemical Society}, 2014, \textbf{136}, 8508--8511\relax
\mciteBstWouldAddEndPuncttrue
\mciteSetBstMidEndSepPunct{\mcitedefaultmidpunct}
{\mcitedefaultendpunct}{\mcitedefaultseppunct}\relax
\EndOfBibitem
\bibitem[Lu and Xiang(2014)]{Lu/Xiang:2014}
X.~Z. Lu and H.~J. Xiang, \emph{Phys. Rev. B}, 2014, \textbf{90}, 104409\relax
\mciteBstWouldAddEndPuncttrue
\mciteSetBstMidEndSepPunct{\mcitedefaultmidpunct}
{\mcitedefaultendpunct}{\mcitedefaultseppunct}\relax
\EndOfBibitem
\bibitem[Kunz and Brown(1995)]{Kunz/Brown:1995}
M.~Kunz and I.~D. Brown, \emph{Journal of Solid State Chemistry}, 1995,
  \textbf{115}, 395--406\relax
\mciteBstWouldAddEndPuncttrue
\mciteSetBstMidEndSepPunct{\mcitedefaultmidpunct}
{\mcitedefaultendpunct}{\mcitedefaultseppunct}\relax
\EndOfBibitem
\bibitem[Halasyamani and Poeppelmeier(1998)]{Halasyamani/Poeppelmeier:1998}
P.~S. Halasyamani and K.~R. Poeppelmeier, \emph{Chem. Mater.}, 1998,
  \textbf{10}, 2753--2769\relax
\mciteBstWouldAddEndPuncttrue
\mciteSetBstMidEndSepPunct{\mcitedefaultmidpunct}
{\mcitedefaultendpunct}{\mcitedefaultseppunct}\relax
\EndOfBibitem
\bibitem[Bersuker(1995)]{Bersuker:1995}
I.~B. Bersuker, \emph{Ferroelectrics}, 1995, \textbf{164}, 75--100\relax
\mciteBstWouldAddEndPuncttrue
\mciteSetBstMidEndSepPunct{\mcitedefaultmidpunct}
{\mcitedefaultendpunct}{\mcitedefaultseppunct}\relax
\EndOfBibitem
\bibitem[Burdett(1981)]{Burdett:1981}
J.~K. Burdett, \emph{Inorganic~Chemistry}, 1981, \textbf{20}, 1959--1962\relax
\mciteBstWouldAddEndPuncttrue
\mciteSetBstMidEndSepPunct{\mcitedefaultmidpunct}
{\mcitedefaultendpunct}{\mcitedefaultseppunct}\relax
\EndOfBibitem
\bibitem[Barone \emph{et~al.}(2011)Barone, Kanungo, Picozzi, and
  Saha-Dasgupta]{Barone/Picozzi_CaMnO3:2011}
P.~Barone, S.~Kanungo, S.~Picozzi and T.~Saha-Dasgupta, \emph{Phys. Rev. B},
  2011, \textbf{84}, 134101\relax
\mciteBstWouldAddEndPuncttrue
\mciteSetBstMidEndSepPunct{\mcitedefaultmidpunct}
{\mcitedefaultendpunct}{\mcitedefaultseppunct}\relax
\EndOfBibitem
\bibitem[Bersuker(2012)]{Bersuker:2012}
I.~B. Bersuker, \emph{Phys. Rev. Lett.}, 2012, \textbf{108},
  137202--1--137202--5\relax
\mciteBstWouldAddEndPuncttrue
\mciteSetBstMidEndSepPunct{\mcitedefaultmidpunct}
{\mcitedefaultendpunct}{\mcitedefaultseppunct}\relax
\EndOfBibitem
\bibitem[Bousquet \emph{et~al.}(2014)Bousquet, Dawber, Stucki, Lichtensteiger,
  Hermet, Gariglio, Triscone, and Ghosez]{Bousquet_etal:2008}
E.~Bousquet, M.~Dawber, N.~Stucki, C.~Lichtensteiger, P.~Hermet, S.~Gariglio,
  J.-M. Triscone and P.~Ghosez, \emph{Phys. Rev. B}, 2014, \textbf{90},
  104409\relax
\mciteBstWouldAddEndPuncttrue
\mciteSetBstMidEndSepPunct{\mcitedefaultmidpunct}
{\mcitedefaultendpunct}{\mcitedefaultseppunct}\relax
\EndOfBibitem
\bibitem[Benedek and Fennie(2011)]{Benedek/Fennie:2011}
N.~A. Benedek and C.~J. Fennie, \emph{Phys. Rev. Lett.}, 2011, \textbf{106},
  107204\relax
\mciteBstWouldAddEndPuncttrue
\mciteSetBstMidEndSepPunct{\mcitedefaultmidpunct}
{\mcitedefaultendpunct}{\mcitedefaultseppunct}\relax
\EndOfBibitem
\bibitem[Martin \emph{et~al.}(2010)Martin, Chu, and
  Ramesh]{Martin/Chu/Ramesh:2010}
L.~W. Martin, Y.-H. Chu and R.~Ramesh, \emph{Materials Science and Engineering
  R}, 2010, \textbf{68}, 89--133\relax
\mciteBstWouldAddEndPuncttrue
\mciteSetBstMidEndSepPunct{\mcitedefaultmidpunct}
{\mcitedefaultendpunct}{\mcitedefaultseppunct}\relax
\EndOfBibitem
\bibitem[Perez-Mato \emph{et~al.}(2004)Perez-Mato, Aroyo, Garcia, Blaha,
  Schwarz, Schweifer, and Parlinski]{PerezMatoHIF:2004}
J.~M. Perez-Mato, M.~Aroyo, A.~Garcia, P.~Blaha, K.~Schwarz, J.~Schweifer and
  K.~Parlinski, \emph{Phys. Rev. B}, 2004, \textbf{70}, 214111\relax
\mciteBstWouldAddEndPuncttrue
\mciteSetBstMidEndSepPunct{\mcitedefaultmidpunct}
{\mcitedefaultendpunct}{\mcitedefaultseppunct}\relax
\EndOfBibitem
\bibitem[Glazer(1972)]{Glazer:1972}
A.~M. Glazer, \emph{Acta Cryst.~B}, 1972, \textbf{28}, 3384--3392\relax
\mciteBstWouldAddEndPuncttrue
\mciteSetBstMidEndSepPunct{\mcitedefaultmidpunct}
{\mcitedefaultendpunct}{\mcitedefaultseppunct}\relax
\EndOfBibitem
\bibitem[Woodward(1997)]{Woodward1:1997}
P.~M. Woodward, \emph{Acta Cryst. B}, 1997, \textbf{53}, 32--43\relax
\mciteBstWouldAddEndPuncttrue
\mciteSetBstMidEndSepPunct{\mcitedefaultmidpunct}
{\mcitedefaultendpunct}{\mcitedefaultseppunct}\relax
\EndOfBibitem
\bibitem[Woodward(1997)]{Woodward2:1997}
P.~M. Woodward, \emph{Acta Cryst. B}, 1997, \textbf{53}, 44--66\relax
\mciteBstWouldAddEndPuncttrue
\mciteSetBstMidEndSepPunct{\mcitedefaultmidpunct}
{\mcitedefaultendpunct}{\mcitedefaultseppunct}\relax
\EndOfBibitem
\bibitem[Akamatsu \emph{et~al.}(2014)Akamatsu, Fujita, Kuge, Sen~Gupta, Togo,
  Lei, Xue, Stone, Rondinelli, Chen, Tanaka, Gopalan, and
  Tanaka]{PhysRevLett.112.187602}
H.~Akamatsu, K.~Fujita, T.~Kuge, A.~Sen~Gupta, A.~Togo, S.~Lei, F.~Xue,
  G.~Stone, J.~M. Rondinelli, L.-Q. Chen, I.~Tanaka, V.~Gopalan and K.~Tanaka,
  \emph{Phys. Rev. Lett.}, 2014, \textbf{112}, 187602\relax
\mciteBstWouldAddEndPuncttrue
\mciteSetBstMidEndSepPunct{\mcitedefaultmidpunct}
{\mcitedefaultendpunct}{\mcitedefaultseppunct}\relax
\EndOfBibitem
\bibitem[Benedek(2014)]{Benedek:2014}
N.~Benedek, \emph{Inorg. Chem.}, 2014, \textbf{53}, 3769--3777\relax
\mciteBstWouldAddEndPuncttrue
\mciteSetBstMidEndSepPunct{\mcitedefaultmidpunct}
{\mcitedefaultendpunct}{\mcitedefaultseppunct}\relax
\EndOfBibitem
\bibitem[Bousquet \emph{et~al.}(2008)Bousquet, Dawber, Stucki, Lichtensteiger,
  Hermet, Gariglio, Triscone, and Ghosez]{Bousquet/Ghosez_et_al:2008}
E.~Bousquet, M.~Dawber, N.~Stucki, C.~Lichtensteiger, P.~Hermet, S.~Gariglio,
  J.-M. Triscone and P.~Ghosez, \emph{Nature}, 2008, \textbf{452},
  732--736\relax
\mciteBstWouldAddEndPuncttrue
\mciteSetBstMidEndSepPunct{\mcitedefaultmidpunct}
{\mcitedefaultendpunct}{\mcitedefaultseppunct}\relax
\EndOfBibitem
\bibitem[Zhao \emph{et~al.}(2014)Zhao, \'I\~niguez, Ren, Chen, and
  Bellaiche]{PhysRevB.89.174101}
H.~J. Zhao, J.~\'I\~niguez, W.~Ren, X.~M. Chen and L.~Bellaiche, \emph{Phys.
  Rev. B}, 2014, \textbf{89}, 174101\relax
\mciteBstWouldAddEndPuncttrue
\mciteSetBstMidEndSepPunct{\mcitedefaultmidpunct}
{\mcitedefaultendpunct}{\mcitedefaultseppunct}\relax
\EndOfBibitem
\bibitem[Zanolli \emph{et~al.}(2013)Zanolli, Wojde\l, \'{I}\~{n}iguez, and
  Ghosez]{IniguezGhosez_etal:2013}
Z.~Zanolli, J.~C. Wojde\l, J.~\'{I}\~{n}iguez and P.~Ghosez, \emph{Phys. Rev.
  B}, 2013, \textbf{88}, 060102\relax
\mciteBstWouldAddEndPuncttrue
\mciteSetBstMidEndSepPunct{\mcitedefaultmidpunct}
{\mcitedefaultendpunct}{\mcitedefaultseppunct}\relax
\EndOfBibitem
\bibitem[Stroppa \emph{et~al.}(2011)Stroppa, Prashant, Barone, Marsman,
  Perez-Mato, Cheetham, Kroto, and Picozzi]{Stroppa_CuMOF:2011}
A.~Stroppa, J.~Prashant, P.~Barone, M.~Marsman, J.~M. Perez-Mato, A.~K.
  Cheetham, H.~W. Kroto and S.~Picozzi, \emph{Angew. Chem. Int. Ed.}, 2011,
  \textbf{50}, 5847--5850\relax
\mciteBstWouldAddEndPuncttrue
\mciteSetBstMidEndSepPunct{\mcitedefaultmidpunct}
{\mcitedefaultendpunct}{\mcitedefaultseppunct}\relax
\EndOfBibitem
\bibitem[Stroppa \emph{et~al.}(2013)Stroppa, Barone, Jain, Perez-Mato, and
  Picozzi]{StroppaMOF:2013}
A.~Stroppa, P.~Barone, P.~Jain, J.~M. Perez-Mato and S.~Picozzi, \emph{Adv.
  Mater.}, 2013, \textbf{25}, 2284--2290\relax
\mciteBstWouldAddEndPuncttrue
\mciteSetBstMidEndSepPunct{\mcitedefaultmidpunct}
{\mcitedefaultendpunct}{\mcitedefaultseppunct}\relax
\EndOfBibitem
\bibitem[Picozzi and Stroppa(2012)]{Picozzi/Stroppa_MF:2012}
S.~Picozzi and A.~Stroppa, \emph{Eur. Phys. J. B}, 2012, \textbf{85}, 240\relax
\mciteBstWouldAddEndPuncttrue
\mciteSetBstMidEndSepPunct{\mcitedefaultmidpunct}
{\mcitedefaultendpunct}{\mcitedefaultseppunct}\relax
\EndOfBibitem
\bibitem[Rondinelli and Fennie(2012)]{Rondinelli/Fennie:2012}
J.~M. Rondinelli and C.~J. Fennie, \emph{Adv. Mater.}, 2012, \textbf{24},
  1961--1968\relax
\mciteBstWouldAddEndPuncttrue
\mciteSetBstMidEndSepPunct{\mcitedefaultmidpunct}
{\mcitedefaultendpunct}{\mcitedefaultseppunct}\relax
\EndOfBibitem
\bibitem[Mulder \emph{et~al.}(2013)Mulder, Benedek, Rondinelli, and
  Fennie]{Mulder/Rondinelli/Fennie:2013}
A.~T. Mulder, N.~A. Benedek, J.~M. Rondinelli and C.~J. Fennie, \emph{Adv.
  Func. Mater.}, 2013,  4810--4820\relax
\mciteBstWouldAddEndPuncttrue
\mciteSetBstMidEndSepPunct{\mcitedefaultmidpunct}
{\mcitedefaultendpunct}{\mcitedefaultseppunct}\relax
\EndOfBibitem
\bibitem[Young and Rondinelli(2013)]{Young/Rondinelli:2013}
J.~Young and J.~M. Rondinelli, \emph{Chem. Mater.}, 2013, \textbf{25},
  4545--4550\relax
\mciteBstWouldAddEndPuncttrue
\mciteSetBstMidEndSepPunct{\mcitedefaultmidpunct}
{\mcitedefaultendpunct}{\mcitedefaultseppunct}\relax
\EndOfBibitem
\bibitem[Benedek \emph{et~al.}(2012)Benedek, Mulder, and
  Fennie]{Benedek/Mulder/Fennie:2012}
N.~Benedek, A.~T. Mulder and C.~J. Fennie, \emph{Journal of Solid State
  Chemistry}, 2012, \textbf{195}, 11--20\relax
\mciteBstWouldAddEndPuncttrue
\mciteSetBstMidEndSepPunct{\mcitedefaultmidpunct}
{\mcitedefaultendpunct}{\mcitedefaultseppunct}\relax
\EndOfBibitem
\bibitem[Bellaiche and \'{I}\~{n}iguez(2013)]{Bellaiche/Iniguez:2013}
L.~Bellaiche and J.~\'{I}\~{n}iguez, \emph{Phys. Rev. B}, 2013, \textbf{88},
  014104\relax
\mciteBstWouldAddEndPuncttrue
\mciteSetBstMidEndSepPunct{\mcitedefaultmidpunct}
{\mcitedefaultendpunct}{\mcitedefaultseppunct}\relax
\EndOfBibitem
\bibitem[Zhao \emph{et~al.}(2014)Zhao, \'{I}\~{n}iguez, Ren, M., and
  Bellaiche]{Bellaiche/Iniguez_etal:2014}
H.~J. Zhao, J.~\'{I}\~{n}iguez, W.~Ren, C.~X. M. and L.~Bellaiche, \emph{Phys.
  Rev. B}, 2014, \textbf{89}, 174101\relax
\mciteBstWouldAddEndPuncttrue
\mciteSetBstMidEndSepPunct{\mcitedefaultmidpunct}
{\mcitedefaultendpunct}{\mcitedefaultseppunct}\relax
\EndOfBibitem
\bibitem[King and Woodward(2010)]{Graham/Woodward:2010}
G.~King and P.~M. Woodward, \emph{J. Mater. Chem.}, 2010, \textbf{20},
  5785--5796\relax
\mciteBstWouldAddEndPuncttrue
\mciteSetBstMidEndSepPunct{\mcitedefaultmidpunct}
{\mcitedefaultendpunct}{\mcitedefaultseppunct}\relax
\EndOfBibitem
\bibitem[Lopez \emph{et~al.}(1994)Lopez, Veiga, and
  Pico]{Lopez/Veiga/Pico:1994}
M.~L. Lopez, M.~L. Veiga and C.~Pico, \emph{J. Mater. Chem.}, 1994, \textbf{4},
  547--550\relax
\mciteBstWouldAddEndPuncttrue
\mciteSetBstMidEndSepPunct{\mcitedefaultmidpunct}
{\mcitedefaultendpunct}{\mcitedefaultseppunct}\relax
\EndOfBibitem
\bibitem[King \emph{et~al.}(2009)King, Wayman, and
  Woodward]{King/WaymanWoodward:2009}
G.~King, L.~M. Wayman and P.~M. Woodward, \emph{Journal of Solid State
  Chemistry}, 2009, \textbf{182}, 1319--1325\relax
\mciteBstWouldAddEndPuncttrue
\mciteSetBstMidEndSepPunct{\mcitedefaultmidpunct}
{\mcitedefaultendpunct}{\mcitedefaultseppunct}\relax
\EndOfBibitem
\bibitem[Knapp and Woodward(2006)]{Knapp/Woodward:2006}
M.~C. Knapp and P.~M. Woodward, \emph{Journal of Solid State Chemistry}, 2006,
  \textbf{179}, 1076--1085\relax
\mciteBstWouldAddEndPuncttrue
\mciteSetBstMidEndSepPunct{\mcitedefaultmidpunct}
{\mcitedefaultendpunct}{\mcitedefaultseppunct}\relax
\EndOfBibitem
\bibitem[King \emph{et~al.}(2007)King, Thimmaiah, Dwivedi, and
  Woodward]{King/Woodward_etal:2007}
G.~King, S.~Thimmaiah, A.~Dwivedi and P.~M. Woodward, \emph{Chem. Mater.},
  2007, \textbf{19}, 6451--6458\relax
\mciteBstWouldAddEndPuncttrue
\mciteSetBstMidEndSepPunct{\mcitedefaultmidpunct}
{\mcitedefaultendpunct}{\mcitedefaultseppunct}\relax
\EndOfBibitem
\bibitem[Arillo \emph{et~al.}(1997)Arillo, Gomez, Lopez, Pico, and
  Veiga]{Arillo/Gomez_etal:1997}
M.~A. Arillo, J.~Gomez, M.~L. Lopez, C.~Pico and M.~L. Veiga, \emph{Solid State
  Ionics}, 1997, \textbf{95}, 241--248\relax
\mciteBstWouldAddEndPuncttrue
\mciteSetBstMidEndSepPunct{\mcitedefaultmidpunct}
{\mcitedefaultendpunct}{\mcitedefaultseppunct}\relax
\EndOfBibitem
\bibitem[Arillo \emph{et~al.}(1997)Arillo, Gomez, Lopez, Pico, and
  Veiga]{Arillo/Gomez_etal_2:1997}
M.~A. Arillo, J.~Gomez, M.~L. Lopez, C.~Pico and M.~L. Veiga, \emph{J. Mater.
  Chem.}, 1997, \textbf{7}, 801--806\relax
\mciteBstWouldAddEndPuncttrue
\mciteSetBstMidEndSepPunct{\mcitedefaultmidpunct}
{\mcitedefaultendpunct}{\mcitedefaultseppunct}\relax
\EndOfBibitem
\bibitem[King \emph{et~al.}(2009)King, Wills, and
  Woodward]{King/Wills/Woodward:2009}
G.~King, A.~S. Wills and P.~M. Woodward, \emph{Phys. Rev. B}, 2009,
  \textbf{79}, 224428\relax
\mciteBstWouldAddEndPuncttrue
\mciteSetBstMidEndSepPunct{\mcitedefaultmidpunct}
{\mcitedefaultendpunct}{\mcitedefaultseppunct}\relax
\EndOfBibitem
\bibitem[Fukushima \emph{et~al.}(2011)Fukushima, Stroppa, Picozzi, and
  Perez-Mato]{Fukushima_etal:2011}
T.~Fukushima, A.~Stroppa, S.~Picozzi and J.~M. Perez-Mato, \emph{Phys. Chem.
  Chem. Phys.}, 2011, \textbf{13}, 12186--12190\relax
\mciteBstWouldAddEndPuncttrue
\mciteSetBstMidEndSepPunct{\mcitedefaultmidpunct}
{\mcitedefaultendpunct}{\mcitedefaultseppunct}\relax
\EndOfBibitem
\bibitem[De \emph{et~al.}(2014)De, Kim, Kim, and Sundaresan]{De_Kim_etal:2014}
C.~De, T.~H. Kim, K.~H. Kim and A.~Sundaresan, \emph{Phys. Chem. Chem. Phys.},
  2014, \textbf{16}, 5407--5411\relax
\mciteBstWouldAddEndPuncttrue
\mciteSetBstMidEndSepPunct{\mcitedefaultmidpunct}
{\mcitedefaultendpunct}{\mcitedefaultseppunct}\relax
\EndOfBibitem
\bibitem[Hohenberg and Kohn(1964)]{Hohenberg/Kohn:1964}
P.~Hohenberg and W.~Kohn, \emph{Physical Review}, 1964, \textbf{136},
  B864--B871\relax
\mciteBstWouldAddEndPuncttrue
\mciteSetBstMidEndSepPunct{\mcitedefaultmidpunct}
{\mcitedefaultendpunct}{\mcitedefaultseppunct}\relax
\EndOfBibitem
\bibitem[Kresse and Hafner(1993)]{Kresse/Hafner:1993}
G.~Kresse and J.~Hafner, \emph{Phys. Rev. B}, 1993, \textbf{47}, 558--561\relax
\mciteBstWouldAddEndPuncttrue
\mciteSetBstMidEndSepPunct{\mcitedefaultmidpunct}
{\mcitedefaultendpunct}{\mcitedefaultseppunct}\relax
\EndOfBibitem
\bibitem[Kresse and Furthm\"uller(1996)]{Kresse/Hafner:1996}
G.~Kresse and J.~Furthm\"uller, \emph{Comput. Mat. Sci.}, 1996, \textbf{6},
  15--50\relax
\mciteBstWouldAddEndPuncttrue
\mciteSetBstMidEndSepPunct{\mcitedefaultmidpunct}
{\mcitedefaultendpunct}{\mcitedefaultseppunct}\relax
\EndOfBibitem
\bibitem[Bl\"ochl(1994)]{Blochl:1994}
P.~E. Bl\"ochl, \emph{Phys. Rev. B}, 1994, \textbf{50}, 17953--17979\relax
\mciteBstWouldAddEndPuncttrue
\mciteSetBstMidEndSepPunct{\mcitedefaultmidpunct}
{\mcitedefaultendpunct}{\mcitedefaultseppunct}\relax
\EndOfBibitem
\bibitem[Perdew \emph{et~al.}(1996)Perdew, Burke, and Ernzerhof]{PBE1:1996}
J.~P. Perdew, K.~Burke and M.~Ernzerhof, \emph{Phys. Rev. Lett.}, 1996,
  \textbf{77}, 3865\relax
\mciteBstWouldAddEndPuncttrue
\mciteSetBstMidEndSepPunct{\mcitedefaultmidpunct}
{\mcitedefaultendpunct}{\mcitedefaultseppunct}\relax
\EndOfBibitem
\bibitem[Perdew \emph{et~al.}(1997)Perdew, Burke, and Ernzerhof]{PBE2:1997}
J.~P. Perdew, K.~Burke and M.~Ernzerhof, \emph{Phys. Rev. Lett.}, 1997,
  \textbf{78}, 1396\relax
\mciteBstWouldAddEndPuncttrue
\mciteSetBstMidEndSepPunct{\mcitedefaultmidpunct}
{\mcitedefaultendpunct}{\mcitedefaultseppunct}\relax
\EndOfBibitem
\bibitem[Monkhorst and Pack(1976)]{Monkhorst/Pack:1976}
H.~J. Monkhorst and J.~D. Pack, \emph{Phys. Rev. B}, 1976, \textbf{13},
  5188--5192\relax
\mciteBstWouldAddEndPuncttrue
\mciteSetBstMidEndSepPunct{\mcitedefaultmidpunct}
{\mcitedefaultendpunct}{\mcitedefaultseppunct}\relax
\EndOfBibitem
\bibitem[Dudarev \emph{et~al.}(1998)Dudarev, Botton, Savrasov, Humphreys, and
  Sutton]{Dudarev:1998}
S.~L. Dudarev, G.~A. Botton, S.~Y. Savrasov, C.~J. Humphreys and A.~P. Sutton,
  \emph{Phys. Rev. B}, 1998, \textbf{57}, 1505\relax
\mciteBstWouldAddEndPuncttrue
\mciteSetBstMidEndSepPunct{\mcitedefaultmidpunct}
{\mcitedefaultendpunct}{\mcitedefaultseppunct}\relax
\EndOfBibitem
\bibitem[King-Smith and Vanderbilt(1993)]{King-Smith/Vanderbilt:1993}
R.~D. King-Smith and D.~Vanderbilt, \emph{Phys. Rev. B}, 1993, \textbf{47},
  R1651--R1654\relax
\mciteBstWouldAddEndPuncttrue
\mciteSetBstMidEndSepPunct{\mcitedefaultmidpunct}
{\mcitedefaultendpunct}{\mcitedefaultseppunct}\relax
\EndOfBibitem
\bibitem[Resta(1994)]{RevModPhys.66.899}
R.~Resta, \emph{Reviews of Modern Physics}, 1994, \textbf{66}, 899--915\relax
\mciteBstWouldAddEndPuncttrue
\mciteSetBstMidEndSepPunct{\mcitedefaultmidpunct}
{\mcitedefaultendpunct}{\mcitedefaultseppunct}\relax
\EndOfBibitem
\bibitem[Campbell \emph{et~al.}(2006)Campbell, Stokes, Tanner, and
  Hatch]{ISODISTORT}
B.~J. Campbell, H.~T. Stokes, D.~E. Tanner and D.~M. Hatch, \emph{J. Appl.
  Crystallogr.}, 2006, \textbf{39}, 607--614\relax
\mciteBstWouldAddEndPuncttrue
\mciteSetBstMidEndSepPunct{\mcitedefaultmidpunct}
{\mcitedefaultendpunct}{\mcitedefaultseppunct}\relax
\EndOfBibitem
\bibitem[Momma and Izumi(2008)]{VESTA}
K.~Momma and F.~Izumi, \emph{J. Appl. Crystallogr.}, 2008, \textbf{41},
  653--658\relax
\mciteBstWouldAddEndPuncttrue
\mciteSetBstMidEndSepPunct{\mcitedefaultmidpunct}
{\mcitedefaultendpunct}{\mcitedefaultseppunct}\relax
\EndOfBibitem
\bibitem[Garcia-Martin \emph{et~al.}(2011)Garcia-Martin, King, Urones-Garrote,
  N\'{e}bert, and Woodward]{GarciaMartin_etal:2011}
S.~Garcia-Martin, G.~King, E.~Urones-Garrote, G.~N\'{e}bert and P.~M. Woodward,
  \emph{Chem. Mater.}, 2011, \textbf{23}, 163--170\relax
\mciteBstWouldAddEndPuncttrue
\mciteSetBstMidEndSepPunct{\mcitedefaultmidpunct}
{\mcitedefaultendpunct}{\mcitedefaultseppunct}\relax
\EndOfBibitem
\bibitem[Altermatt and Brown(1985)]{Brown/Altermatt:1985}
D.~Altermatt and I.~D. Brown, \emph{Acta Cryst. B}, 1985, \textbf{41},
  240--244\relax
\mciteBstWouldAddEndPuncttrue
\mciteSetBstMidEndSepPunct{\mcitedefaultmidpunct}
{\mcitedefaultendpunct}{\mcitedefaultseppunct}\relax
\EndOfBibitem
\bibitem[Cammarata and Rondinelli(2014)]{Cammarata/Rondinelli:2014}
A.~Cammarata and J.~M. Rondinelli, \emph{J.~Chem. Phys.}, 2014, \textbf{141},
  114704\relax
\mciteBstWouldAddEndPuncttrue
\mciteSetBstMidEndSepPunct{\mcitedefaultmidpunct}
{\mcitedefaultendpunct}{\mcitedefaultseppunct}\relax
\EndOfBibitem
\bibitem[Goldschmidt(1926)]{Goldschmidt:1926}
V.~M. Goldschmidt, \emph{Naturwissenschaften}, 1926, \textbf{14},
  477--485\relax
\mciteBstWouldAddEndPuncttrue
\mciteSetBstMidEndSepPunct{\mcitedefaultmidpunct}
{\mcitedefaultendpunct}{\mcitedefaultseppunct}\relax
\EndOfBibitem
\bibitem[Shannon(1976)]{Shannon:1976}
R.~D. Shannon, \emph{Acta Cryst. A}, 1976, \textbf{32}, 751--767\relax
\mciteBstWouldAddEndPuncttrue
\mciteSetBstMidEndSepPunct{\mcitedefaultmidpunct}
{\mcitedefaultendpunct}{\mcitedefaultseppunct}\relax
\EndOfBibitem
\end{mcitethebibliography}
\bibliographystyle{rsc} 
}

\end{document}